\documentclass{article}
\usepackage{LaThuileFPSpro}

\begin{document}

\title{ 
  PROSPECTS IN CP VIOLATION MEASUREMENTS AT THE TEVATRON COLLIDER
  }
\author{
  Diego Tonelli        \\
(representing the CDF II and the D\O~Collaborations)\\
  {\em Scuola Normale Superiore and Istituto Nazionale di Fisica
 Nucleare of Pisa}\\ {\em Edificio C - Polo Fibonacci, Via Buonarroti
 2, 56100 Pisa - Italy} \\ {\em E-mail: tonel@fnal.gov}
    }
\maketitle
   
\baselineskip=11.6pt

\begin{abstract}
The Fermilab Tevatron Collider is currently the most copious source
of $b$-hadrons, thanks to the large $b\bar{b}$ production cross-section in
 1.96 TeV $p\bar{p}$ collisions. Recent detector upgrades allow for a wide
 range of \textsf{CP} violation and flavor-mixing measurements that 
are fully competitive (\emph{direct} asymmetries in self-tagging modes) 
or complementary  (asymmetries of $B_s$ and  $b$-baryons decays) 
with $B$-factories. In this  paper we review some recent \textsf{CP}
 violation results from the D\O~and  CDF II Collaborations and we discuss the prospects for future measurements.
\end{abstract}

\newpage
\section{Introduction}
\label{sec:intro}
Although several  substantial improvements have been achieved
in understanding the \textsf{CP}  violation mechanism,
 we are left with several open  questions, and an experimental effort
 is necessary to increase our comprehension.
 The $b$ sector offers many  interesting processes in which large
 \textsf{CP} violating effects are possible, some 
of which have clean theoretical interpretation.
 Traditionally, $b$ physics has been the
 domain of $e^+e^-$  machines operating at the $\Upsilon$(4S) resonance
 ($B$-factories) or at the $Z^0$ pole. In particular, the recent successful
 turn on of the Babar  and Belle  experiments  provided an impressive
 precision in many experimental 
measurements. Nevertheless, already in 1990 the UA1 Collaboration  proved 
that $b$ physics is accessible in an hadron  collider environment\cite{UA1}, 
and in 1992 the  CDF Collaboration published  the first signal 
of fully reconstructed $B$ meson decays\cite{CDF1}. 
Afterward, the CDF and D\O~Collaborations 
pursued a  successful $b$ physics program during the 1992-1996 data 
taking period (Run I).\par Since then the experimental techniques improved
 significantly,  with the development of more precise silicon microvertex
 detectors and online triggering of tracks from long-lived particles.
 The Tevatron $p\overline{p}$ Collider, along with the upgraded Collider
 Detector at Fermilab (CDF II) and the D\O~detector, offer today a unique  
opportunity to study $b$ physics. 
 Tevatron results are fully competitive  with $B$-factories and,
 in many cases, the Tevatron measurements are complementary to those 
performed at the $B$-factories.\par
We summarize here some of the recent experimental progress in the
 measurements related to \textsf{CP} violation and flavor-mixing in the 
\emph{beauty} and \emph{charmed} sectors at the Tevatron. Charge-conjugate 
modes are implied throughout all this paper unless otherwise specified.

\section{The Upgraded Tevatron Accelerator}

\label{sec:tev}
The Tevatron accelerator complex has undergone an extensive upgrade since 
the end of Run I. The major improvement is the replacement of the  
final injection stage with the Main Injector, a new  150 GeV proton 
 ring which provides more efficient injection and higher proton intensity onto
 the anti-proton production target\footnote{Actually, in $\bar{p}$ production
 mode, the Main Injector accelerates protons to 120 GeV.}. In Run II (mid-2001 $-$ now)
 the Tevatron accelerates  36 bunches of protons  against 36 bunches of
  anti-protons producing one
 collision every 396 ns at 1.96 TeV centre-of-mass energy.
 The current centre-of-mass  energy, higher than in  Run I 
(1.8 $\rightarrow$ 1.96 TeV),  increases by $\sim$ 10 $-$ 30\% 
the production cross-section for heavy flavors. The  luminous
 regions along the  beamline are about 30 cm long (RMS), requiring
 properly designed  silicon micro-vertex detectors to provide good
 coverage.  The transverse beam-width at the collision points is about
 $25-30~\mu$m (RMS). This  is sufficiently small compared to  the typical 
transverse decay length\footnote{$L_{xy}=\beta_T\gamma c \tau$ where
 $\beta_T\gamma$ is the Lorentz boost projected onto the plane perpendicular
 to the beam line ($\simeq$ 0.5 $-$ 2 at Tevatron) and $c\tau$ is 
the proper decay length of the $b$-hadron ($\simeq$ 450 $\mu$m).} of
 $b$-hadrons, $L_{xy}\simeq 450~\mu$m, 
 to allow a clean separation of secondary from primary vertices. 
The instantaneous luminosity ($\mathcal{L}_{inst}$)
 has been rising steadily since the beginning
 of Run II. A factor of two increase has been achieved just
 during the last  year (March 2003 $-$ March 2004) up to a record of
$\mathcal{L}_{inst}\simeq 6.7\times 10^{31}$cm$^{-2}$s$^{-1}$.
 The  machine regularly exceeds 
$\mathcal{L}_{inst}=5\times 10^{31}$cm$^{-2}$s$^{-1}$ typically delivering
 data corresponding to 10 pb$^{-1}$/week  of integrated luminosity per
 experiment. At such luminosities on average 1.5 interactions per 
bunch-crossing occur. The total integrated luminosity  delivered is around 
400 pb$^{-1}$ per experiment but the data taken during the
 first year were used for machine and detector commissioning. 
Afterward, about  290 pb$^{-1}$ of physics quality data have 
been recorded on tape  by each experiment with typical data-taking
 efficiencies in excess 
of 85\%.  The CDF II results shown in this document used
 up to 190 pb$^{-1}$ of data; the D\O~results used up  to 250 pb$^{-1}$
 of data. 
\section{Overview:  \emph{b} Physics at the Tevatron}
\label{sec:b-phys}
The $b$-hadron phenomenology  in  $p\bar{p}$ collisions at 1.96 TeV offers
 several advantages  with respect to $e^{+}e^{-}$ collisions.
 The Tevatron $b\bar{b}$ production cross-section 
 is very large,  $\mathcal{O}$(100 $\mu$b), compared to the typical $e^+e^-$
 cross-sections at the $\Upsilon$(4S) ($\sim 1$ nb) and $Z^0$ ($\sim 7$ nb)
 resonances.
 Typical production rate  of $b$ quarks at the Tevatron is about 5 kHz and 
they are produced mainly\footnote{The contributions of $p\bar{p}\rightarrow 
\Upsilon(4S)~X$, $\Upsilon(5S)~X \rightarrow[\bar{b}b]~X$, 
$p\bar{p}\rightarrow Z^0X\rightarrow[\bar{b}b]~X$, 
$p\bar{p}\rightarrow W^-X\rightarrow[\bar{c}b]~X$ and
$p\bar{p}\rightarrow W^-X\rightarrow[\bar{t}b]~X$ are neglected.} incoherently
 in $b\bar{b}$ pairs by strong interaction.
 As a consequence, mixing and \textsf{CP} 
violation measurements can be performed 
 by reconstructing a single $b$-hadron in the event, while at $B$-factories
 the flavor of one $B$ meson is determined only after observing 
the decay of the other.
 Moreover, unlike the $B$-factories,  all species of  $b$ hadrons 
are produced at the Tevatron, including  $B_s$, $B_c$ and $b$-baryons 
such as $\Lambda_b$ and $\Xi_b$.\par However the hadronic environment poses several challenges for the $b$-physics.
 The very large $b\bar{b}$ production cross-section is only 
about 1/500$^{th}$ the total inelastic $p\bar{p}$ cross-section,
 $\sigma_{inel}(p\bar{p})\simeq$ 50 mb.
 Moreover, Tevatron events show high ($\approx$ 50) 
 track multiplicities, due to the  fragmentation of the hard interaction 
products, to the underlying  events (\emph{i.e.} hadronized remnants 
of $p$ and $\bar{p}$) and  to pile-up events (multiple collisions per bunch
 crossing). One way to identify $b$-decays in such a complicate environment
is to exploit the relatively long lifetime of $b$-flavors resulting in decay
vertices which are separated by hundreds of microns  from
 the primary $p\bar{p}$ interaction vertices. 
In addition,  the distribution of the transverse
 momentum\footnote{$p_T$ is the particle momentum  component perpendicular
 to the beam line: $p_T = p\cdot\sin(\theta)$ where  $\theta$ is the polar
 angle with respect to the beam axis ($\theta=0 \Rightarrow$ particle 
collinear with the proton direction).}  ($p_T$) of  $b$-hadrons at the 
Tevatron  is a steeply  falling function. Most of the $b$-hadrons
 have very low $p_T$ and  decay into particles which are typically quite soft,
  often having $p_T<1$ GeV/c. This  has two consequences: (1) the need  to select tracks as
 soft as possible conflicts with the limited bandwidth  allowed by the data
 acquisition systems; (2)  since the longitudinal  component  
of $b$-hadrons  momentum is frequently very large, they tend to decay into 
particles  boosted along the beam line, thus  escaping the detector 
acceptance. If one $b$-quark is central 
(pseudorapidity\footnote{The pseudorapidity $\eta$ of a particle  is defined
 as $\eta= -\ln\tan(\theta/2)$.} $|\eta|<$1),
 the other one is central only $\mathcal{O}(10\%)$ of the time.\par
A key ingredient for an effective $b$-hadron selection and 
reconstruction at the Tevatron is, therefore,  an excellent tracking system 
with reconstructing capability extended to low transverse momenta
. The tracking system
 should also provide excellent vertexing performances to 
select long-lived decays
  that are likely to contain heavy flavors. Finally, it is necessary a
 selective trigger and data  acquisition system
 capable to sustain the high rates associated with this 
physics. In the following section we outline some of the relevant aspects of the CDF II and D\O~detectors and triggers.   
\section{The CDF II and D\O~Detectors}
\label{sec:detect}
The CDF II and D\O~detectors are large multipurpose solenoidal magnetic
 spectrometers surrounded by  $4\pi$ calorimetry and muon filters.
 They are axially and azimuthally symmetric around the interaction point 
and their strengths are somewhat complementary to one another. D\O~has a 
excellent tracking acceptance  and very good $e$ and $\mu$ identification
 performance. CDF II features very precise
 tracking  that provides excellent  mass resolution and has strong  particle
 identification capabilities. Additional details of the detectors can be found 
elsewhere\cite{do-tdr}\cite{cdf-tdr}.
\subsection{D\O}
\label{ssec:D0}
The D\O~tracking volume consists of an inner silicon detector surrounded 
by a scintillating fiber tracker both immersed in a  2 T solenoidal field
 with 52 cm total lever arm. The fiber tracker covers the
 region $|\eta|<1.7$, the silicon detector is organized in longitudinal
 barrels interspersed with  azimuthal disks which extend the forward tracking
 to $|\eta|<3$. Tracks with transverse momentum as low as 180 MeV/c are
 reconstructed.  The uranium/liquid-argon calorimeter has very good 
energy resolution  for $e$, $\gamma$ and hadronic jets 
and features pre-shower counters to improve $e/\gamma$ discrimination.
  The muon system covers $|\eta|<2$ for muons with $p_T > 2 - 4.5$ GeV/c.
\subsection{CDF}
\label{ssec:CDF}
 The CDF II tracking system consists of an inner silicon system 
surrounded by a gas-wire drift chamber both immersed in a 1.4 T field with
 135 cm total lever arm. Six to seven double-sided silicon layers, plus
 one single-sided layer, cover the full Tevatron luminous region ($|\eta|<2$)
 and extend radially from 1.6 to 28 cm from the beam line.
 The drift chamber provides 96  
(48 axial plus 48 stereo) samplings of track paths within $|\eta|< 1$. 
Tracking information is integrated with muon (reconstructed at $|\eta|<1.5$) 
and calorimeter ($|\eta|<2$) data for  $\mu$ and  $e/\gamma/$hadron 
identification.  Low-momentum particle identification (PID) is performed
 using  a scintillator-based  Time-of-Flight detector (TOF)  with 110 ps
 resolution  that provides $2\sigma$ K/$\pi$ separation at $p<1.5$ GeV/c.
 The specific ionization information from the drift
 chamber (\emph{dE/dx}) complements the PID with 1.4$\sigma$ K/$\pi$
 separation for tracks with $p>2$ GeV/c.
\section{The CDF II and D\O~Triggers}
\label{sec:trig}
The trigger system is probably the single most important ingredient to pursue
 an effective \emph{b} physics  program at the Tevatron. Both CDF II 
and D\O~have
a multi-stage trigger system organized in three Levels. CDF II has a
 data acquisition system faster than D\O, allowing for a higher Level 1 Accept 
trigger rate, while D\O~has Level 2 Accept rate higher 
than CDF II.\par Past experience from Run I suggests that triggering on 
final states  containing  single or di-leptons is a successful strategy
 to select high  statistics  samples of $b$-hadron decays. Semi-leptonic
 $B\rightarrow l\nu_{l}X$  plus charmonium 
$B\rightarrow J/\psi X\rightarrow [l^+l^-]X$ decays are $\mathcal{O}(20\%)$
 of $B$ meson widths.
Both CDF II and D\O~adapted such a \emph{``conventional''} approach to 
the upgraded detectors. In addition CDF II has a new capability 
to select events based upon track impact parameter\footnote{
The impact parameter of a track $d_0(tr)$ is the minimum distance between 
the track projection and the primary vertex in the plane transverse 
to the beam.}. The following subsections describe the main features of
 the triggers oriented to $b$-physics. 
\subsection{Conventional Triggers}
\label{ssec:ltrig}
Identification of di-muon events down to very low momentum
 is possible, allowing for efficient $J/\psi\rightarrow \mu^+\mu^-$ 
and rare-decays triggers.  Both experiments trigger upon
 $J/\psi$ decays, 15\% of which come from $b$-hadrons, and then fully 
reconstruct several useful decay modes such as
  $B_s\rightarrow J/\psi\phi\rightarrow [\mu^+\mu^-][K^+K^-]$.
 Although CDF II and D\O~results shown here use only di-muon modes,
 future analyses will include data collected through di-electron triggers. 
\\ D\O~has an inclusive muon trigger with 
excellent acceptance, that collects very large samples of semileptonic
 decays.
 The CDF II semileptonic  triggers require an additional displaced  track
 associated with the lepton, providing cleaner samples with
 smaller yields and are described in the next subsection.
 Trigger thresholds are summarized in Table~\ref{tab:trigger}.
\subsection{Triggering on Displaced Tracks}
\label{ssec:SVT}
A revolutionary feature of CDF II is the ability to trigger events
 containing tracks originated in a vertex displaced from the
 primary. These events are enriched in heavy flavor contents, thanks to
 the higher mean-valued lifetimes of $b$-hadrons. The CDF II
Silicon Vertex Trigger (SVT)\cite{SVT} identifies displaced 
tracks by measuring 
their impact parameter with an intrinsic  
resolution\footnote{The intrinsic impact parameter resolution combined 
with the beam-width $\sigma_{beam}\simeq 30$ $\mu$m determines
 the total impact parameter resolution: 
$\sigma_{SVT}(d_0)\oplus\sigma_{beam}\simeq 47$ $\mu$m.}
 $\sigma_{SVT}(d_0)\simeq$ 35 $\mu$m. Such a high accuracy is 
required to discriminate  $b$-decays  from background tracks and
 can be reached only using the silicon information.
 The experimental challenge is to read-out the silicon 
detector  and perform pattern recognition while sustaining the high
 trigger rate. In a typical latency of 25 $\mu$s/event, 
SVT reconstructs with offline-quality 
two-dimensional tracks (in the  plane transverse to the beam) by combining
 the drift chamber information with the silicon hits.\par
 Two trigger strategies exploit displaced tracks. The 
\emph{``lepton + displaced track''} trigger requires a displaced 
 track associated to an electron (or muon) to select very clean
 samples of semileptonic $b$ decays.
 The \emph{``two track''} trigger requires only two displaced tracks and
 selects exclusive non-leptonic  $b$-decays for the first time in an hadronic
 collider. This trigger accumulates large and clean samples of several
 modes relevant for \textsf{CP} violation and  flavor mixing physics
 such as $B^0\rightarrow\pi^+\pi^-$,  $B_s\rightarrow D_s^-\pi^+\rightarrow[\phi\pi^-]\pi^+\rightarrow[[K^+K^-]\pi^-]\pi^+$ and large
 \emph{charmed} samples as well.\par D\O~currently employs
 an impact-parameter-based trigger at Level 3 (software trigger) and is 
commissioning  a hardware silicon-based  track trigger at Level 2.  
\begin{table}[t]
  \centering
  \caption{ \it CDF II and D\O~trigger thresholds.
    }
  \vskip 0.1 in
  \begin{tabular}{l|l|l}
	
    &  ~~~~~~~~~~~~~~~\scriptsize{CDF ~~~($|\eta|<$1)} & ~~~\scriptsize{D\O~ ~~~($|\eta|<$2)}\\
    \hline
    \scriptsize{inclusive muon}   &   & \scriptsize{$p_T(\mu) >$ 3 GeV/c}\\ \hline
    \scriptsize{di-muon}   & \scriptsize{$p_T(\mu) > 1.5$ GeV/c}  & \scriptsize{$p_T(\mu) > 2.0- 4.5$ GeV/c}\\ \hline
    \scriptsize{e + displ. track}   & \scriptsize{$p_T(e) > 4$ GeV/c} & \\		
      & \scriptsize{$p_T(tr.) >$ 2 GeV/c, $d_0(tr.)>120~\mu$m}  & \\	\hline
    \scriptsize{$\mu$ + displ. track}   & \scriptsize{$p_T(\mu) > 1.5$ GeV/c}  & \\
       & \scriptsize{$p_T(tr.) > 2$ GeV/c, $d_0(tr.)>120~\mu$m} & \\ \hline
    \scriptsize{2 displ. tracks}   & \scriptsize{$p_T(tr.) > 2$ GeV/c} & \\
      &  \scriptsize{$\sum p_T(tr.)> 5.5$ GeV/c, $d_0(tr.)>100~\mu$m}  & \\
    \hline
  \end{tabular}
  \label{tab:trigger}
\end{table}
\section{The Physics Program on \textsf{CP} Violation and Mixing}
\label{sec:phys}
A broad range of competitive measurements on \textsf{CP} violation and flavor-mixing 
physics is  accessible at the Tevatron with the statistics expected before
 the start-up of the Large Hadron Collider ($\mathcal{L}\approx$ 4 fb$^{-1}$
 by the end of 2007). Measurements involving $B_s$ mesons
and $b$-baryons are unique to the Tevatron and play certainly a central 
role in the  physics program of CDF II and D\O. However many measurements 
in the $B_{d(u)}$ (CDF II, D\O) and \emph{charmed} (mainly CDF II) 
sector will  be competitive with the $B$-factories results as well.
 In particular
 the high yields give  some advantage to Tevatron experiments 
in \emph{direct} \textsf{CP} violation measurements with 
self-tagging modes.\par The exclusive opportunity to collect large
 samples of $B_s$ mesons  gives CDF II and D\O~the  privileged possibility
 to study two crucial, and still unknown, parameters of the CKM mechanism: the
 $B_s$ oscillation frequency $\Delta m_s$ and the Bjorken angle 
$\gamma=Arg[-V_{ud}V_{ub}^*/V_{cd}V_{cb}^*]$.  The $B_s$ oscillation frequency allows 
to determine the ratio $\Delta m_s/\Delta m_d$ 
 which is proportional to the length $|\frac{V_{ts}}{V_{td}}|^2$ of 
one side of the main CKM unitarity triangle. On the other hand,  
the simultaneous study of $B^0$ and $B_s$ decays into two charged 
 hadrons ($B_{d(s)}\rightarrow h^{+}h^{'-}$, $h$ being $\pi$ or $K$) 
is  a promising strategy  to extract information on the
  $\gamma$ angle avoiding  the uncertainties from hadronic 
 processes.\par Many other measurements complement the above benchmark
 goals. Both experiments will study the 
 $V_{ts}$ weak phase  $\beta_s= Arg[-V_{ts}V^{*}_{tb}/V_{cs}V^{*}_{cb}]$ using
 $B^0_s\rightarrow J/\psi\phi$ samples. The CDF II displaced track trigger
 provides large hadronic samples where \emph{direct}  \textsf{CP} asymmetries can
 be searched. A few examples are the \emph{charmed} processes
 $D^0\rightarrow K^+K^-/\pi^+\pi^-$, the $B^\pm\rightarrow\phi K^\pm$ decays,
 the  $\Lambda^0_b\rightarrow pK^-, p\pi^-$ modes. 
 In a longer term, CDF II plans 
also to extract additional information on the CKM 
angle $\gamma$ from the $B_{d(s)}\rightarrow D_{(s)}K$ decays.\\
In the following we review the preliminary results of some of 
the mentioned measurements and we discuss the future perspectives. 
\subsection{$B_s$ Flavor Mixing}
\label{ssec:bmix}
In the $K^0$ and $B^0$ systems, particle-antiparticle mixing has been observed
and measured. In particular, $B$-factories experiments
 have significantly improved the world average of $B^0\overline{B^{0}}$ mixing 
frequency up to $\Delta m_d = 0.502 \pm 0.006$ ps$^{-1}$ 
(in terms of a mass difference between the heavy/light eigenstates)\cite{hfag}.
 Mixing proceeds via a  second-order weak transition that involves the
 $V_{td}$ matrix element for  $B^0\overline{B^0}$ mixing, which is replaced
 by $V_{ts}$ in the $B_s\overline{B}_s$ case. Since experimentally 
$\Re(V_{ts})\simeq 0.040 > \Re(V_{td})\simeq 0.007$, we expect the $B_s$
 system oscillate at much higher frequency than the $B^0$ system. To date, 
in fact, $B_s$ oscillations have not yet been resolved. The current 
combined world  limit  sets $\Delta m_s > 14.5$ ps$^{-1}$ 
(at 95\% CL)\cite{hfag} \emph{i.e.} a beam of 
$B_s$ mesons would fully oscillate in less than 1/7$^{th}$ of their lifetime.
 For the next several years, the Tevatron is 
the exclusive laboratory for $B_s$ meson studies, including the search for 
$B_s$ mixing.\par CDF II and D\O~will measure a differential decay rate 
between $N_{mix}$, the number of decays  occurred after mixing, and 
$N_{unmix}$, those occurred without mixing. The decay asymmetry is a
 function of the oscillation frequency:
\begin{equation} 
A_{mix}(t) = \frac{N_{mix}(t)-N_{unmix}(t)}{N_{mix}(t)+N_{unmix}(t)} =
 -\cos(\Delta m_s t) 
\label{eq:bmix1}
\end{equation}
Four ingredients are needed to measure $B_s$ mixing:
\begin{enumerate}
\item {\bf Flavor at the time of production:} it is necessary to know whether the
meson was produced as a $B_s$ or a $\overline{B}_s$.
\item {\bf Flavor at the time of decay:} it is necessary to know whether the
meson was a $B_s$ or a $\overline{B}_s$ when it decayed. This, combined
 with the flavor at time of production, determines whether the meson had
 decayed before\footnote{A meson with the same flavor at time of production
 as at decay could have mixed and mixed back. The time-dependent analysis can
 not discern ``unmixed'' meson from those that underwent one or more complete
 cycles.} or after mixing.
\item {\bf Proper decay time:} it is necessary to know the proper decay time for 
the $B_s$ since we measure the mixing probability as a function of decay time.
Oscillations in the $B_s$ system are too fast to be resolved with
 time-integrated techniques.
\item {\bf Large $B_s$ samples:} the mixing probability should be sampled for 
at least part of the decay spectrum. Since fulfilling each of the previous
 three requirements  reduces the initial available statistics,
 large $B_s$ decays samples are required.
\end{enumerate}
 \subsubsection{\it Initial and Final-State Flavor Tagging}
\label{sssec:flavtag}
Two approaches are adopted for initial-state flavor tagging: the 
 \emph{opposite side} algorithms infer the flavor of the $B_s$ 
from other information in the event, the \emph{same side} algorithms identify
 the $B_s$ flavor by looking at its own fragmentation.\par If a $b\bar{b}$ 
pair is produced in a Tevatron event, about 11\% of the time
 the $\bar{b}(b)$-quark  fragments into a $B_s$($\overline{B}_s$)
 meson that can be selected by the trigger.  The remaining $b$-quark 
(which we refer to as ``the \emph{other} $b$'')
 hadronizes independently\footnote{Here one assumes negligible the color
 correlations between the products of the hard interaction.} 
   into another $b$-hadron.
 If this \emph{other} $b$-hadron enters the detector
acceptance, its  flavor can be measured with several methods 
and it can be used to infer the flavor of the $B_s$ candidate.
 The Soft Lepton Tagging (SLT) exploits the correlation
between the charge of the lepton from  the semileptonic decay of the 
\emph{other} $b$-hadron  and  its flavor 
($b\rightarrow l^-X$ while $\bar{b}\rightarrow l^+X$). The Jet 
Charge algorithm (JeTQ) measures the  momentum-weighted average charge of the
\emph{other} $b$-jet that is correlated with the \emph{other} $b$-quark
 charge. The Opposite Kaon Tagging (OKT) exploits the decay chain  
$b\rightarrow c\rightarrow s$: the charge of the
kaon suggests the flavor of the \emph{other} $b$ ($K^-$ comes likely from 
$\overline{B^0}$). Such techniques suffer from the limited 
acceptance for the \emph{other} hadron and are rather inaccurate.
 SLT tag is wrong if
 the lepton comes from a sequential decay $b\rightarrow c\rightarrow l^+X$.
In case that the \emph{other} $b$ hadron is a neutral meson, mixing can occur 
 before its decay, producing a wrong tag.\par Same Side Tagging (SST) instead
looks at  tracks nearby the triggered $B_s$ meson. 
In particular,
 SST is aimed at exploiting the correlation between the $B_s$ flavor 
and the charge of a near track produced in the fragmentation.
 For a $\bar{b}$ quark to become a $B_s$ meson, for instance, 
it must grab an $s$ quark from the vacuum. 
An accompanying  $\bar{s}$ will be popped from the vacuum  which 
could potentially  fragment into a $K^+$ meson. Alternatively, one could 
exploit $B^{**+}\rightarrow B^0\pi^+$ decays, 
where the pion charge determines the
 $B^0$ flavor. Again this technique is inaccurate, since the
 correlation could be washed out by other fragmentation tracks or the 
charge information could be lost into neutral particles, like $K^0_S$.\par
The performance of initial-state flavor tagging is quoted in terms of 
tagging power: $\epsilon D^2$ where $\epsilon$ is the fraction of times the 
algorithm converged to a  tagging decision, and the dilution $D$ measures the
 probability of a correct tagging. The dilution is 
defined as: $D=(N^{R}-N^{W})/(N^{R}+N^{W})$ where $N_{R}(N_W)$ are the numbers
of right (wrong) tags.  Experimentally, the asymmetry 
of Eq.~(\ref{eq:bmix1}) is reduced  by the effect of incorrect or
 inefficient tagging and becomes:
 $A^{exp}_{mix}(t)=-\sqrt{\epsilon D^2}\cos(\Delta m_st)$.\par
Table~\ref{tab:epsilond2} shows the preliminary tagging performances for
CDF II and D\O.
\begin{table}[t]
  \centering
  \caption{ \it CDF II and D\O~Performance in Flavor Tagging}
  \vskip 0.1 in
  \begin{tabular}{lcc}
      Tagger  & CDF II $\epsilon D^2$ [\%] & D\O~$\epsilon D^2$ [\%]\\
    \hline
Soft Muon & 0.7$\pm$0.1 & 1.6$\pm$1.1\\
Soft Electron & in progress & in progress\\
Jet Charge & 0.42$\pm$0.02& 3.3$\pm$1.7\\
Same Side Pion & 2.4$\pm$1.2 & 5.5$\pm$2.0\\
Same Side Kaon & in progress & -\\
Opposite Side Kaon & in progress & -\\
    \hline
  \end{tabular}
  \label{tab:epsilond2}
\end{table}
D\O~tested the performances of SST, JetQ and SLT  (with muons) using 
$B^+\rightarrow J/\psi K^+\rightarrow [\mu^+\mu^-]K^+$ decays.
 CDF II  used decays collected by its semi-muonic trigger to measure the
 SLT performance. The JetQ performance was measured in both 
semi-leptonic samples. $B^+\rightarrow J/\psi K^+$
 plus $B^+\rightarrow \overline{D^0} \pi^+$ decays were used
to test the Same Side  pion algorithm.
Although the flavor-tagging optimization is still in progress at
 both experiments, the high tracking acceptance and the inclusive
muon trigger seem to provide D\O~an advantage in SLT, JeTQ and pion tagging.
However CDF II, thanks to its PID capabilities, will  exploit kaon tagging.\par
The final-state flavor tagging is straightforward: both experiments collects 
$B_s$ samples with flavor-specific final states such as: $B_s\rightarrow D^-_s\pi^+$ or $B_s\rightarrow D^-_sl^+\nu_l$.
\subsubsection{\it Proper Decay Time}
\label{sssec:ctau}
The mixing frequency is determined by maximizing, with respect to 
$\Delta m_s$, a likelihood function derived from measured and expected 
asymmetries. The height of the maximum likelihood value, 
compared with the second  highest peak  or  some asymptotic value at large 
$\Delta m_s$,  determines  the significance of a mixing observation.
 To a good approximation,
 the  \emph{average} significance is written as\cite{mixsig}:
\begin{equation}
\mathsf{SIG}(\Delta m_s) = \sqrt{\frac{S\epsilon D^2}{2}}e^{\frac{1}{2}\cdot(\Delta m_s\cdot\sigma_{c\tau})^2}\sqrt{\frac{S}{S+B}}
\label{eq:bmix2} 
\end{equation}
where $S~(B)$ are the signal (background) events and $\sigma_{c\tau}$ is the
 average resolution on the measurement of the $B_s$ proper decay length.
 From the exponential  dependence in the above equation it is clear
 the critical role of  $\sigma_{c\tau}$ for the oscillation measurement. Since 
$c\tau=L_{xy}/\beta_T\gamma = L_{xy}\cdot M_B/p_T$, with $p_T$ ($M_B$) the transverse momentum (mass) of the $B_s$, the  
contributions  to $\sigma_{c\tau}$ are:
\begin{equation}
\sigma_{c\tau} = \left(\frac{M_B}{p_T}\right)\sigma_{L_{xy}}\oplus \left(\frac{c\tau}{p_T}\right)\sigma_{p_T} \oplus \left(\frac{L_{xy}}{p_T}\right)\sigma_{M_B}
\label{eq:bmix3} 
\end{equation}
The first term $\sigma_{L_{xy}}$, given the topology and
 kinematics of the decay, depends upon  the tracking and
 vertexing performance of the  detector. The second term $\sigma_{p_T}$ is
 the uncertainty in the time dilation correction. Its contribution is small
 for fully reconstructed decays, where the kinematics is closed and $p_T$
is measured precisely from daughter tracks. In partially reconstructed modes 
(\emph{i.e.} $B_s\rightarrow l^-\nu_l X$),
 the uncertainty on the $B$ meson momentum
contributes significantly ($\mathcal{O}(15\%)$) to the proper time
 uncertainty. The last term\footnote{$\sigma_{M_B}$ is very small compared 
to other uncertainties.} can be neglected in all cases.\par
CDF II estimated $\sigma_{c\tau}\simeq$ 67 fs proper time resolution in
 a sample of  fully  reconstructed $B_s\rightarrow D^-_s\pi^+$ decays basing
on $\sigma_{L_{xy}}\approx 50~\mu$m.
 D\O~performance is expected around $\sigma_{c\tau}\simeq 100$ fs (exclusive) 
and  $\sigma_{c\tau}\simeq 150$ fs (semileptonic). 
\subsubsection{\it $B_s$ Mixing Samples: Semileptonic Decays}
\label{sssec:bssemilept}
If the true value of $\Delta m_s$ is close to the current limit 
($\Delta m_s\sim 14 - 18$ ps$^{-1}$), semileptonic modes will contribute
 to the mixing measurement, since the large event yields somewhat offset
 the poor
 proper time resolution. Otherwise, if $\Delta m_s > 20$ ps$^{-1}$,
 $\sigma_{c\tau}$ becomes the limiting factor and semileptonic 
modes will help mainly for flavor-tagging calibration.\par
\begin{figure}[t]
  \vspace{6.cm}
  \includegraphics{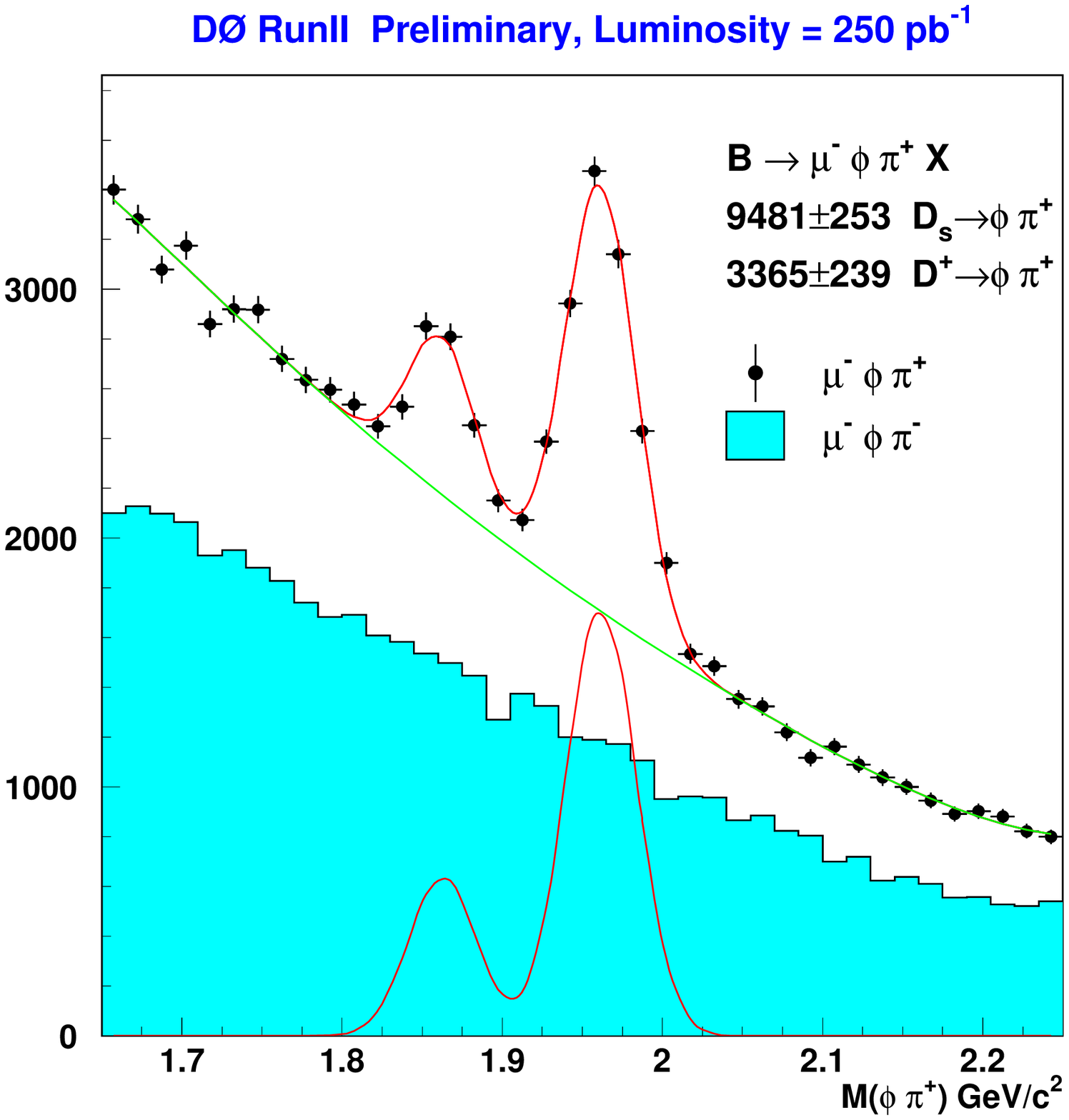}
 \includegraphics{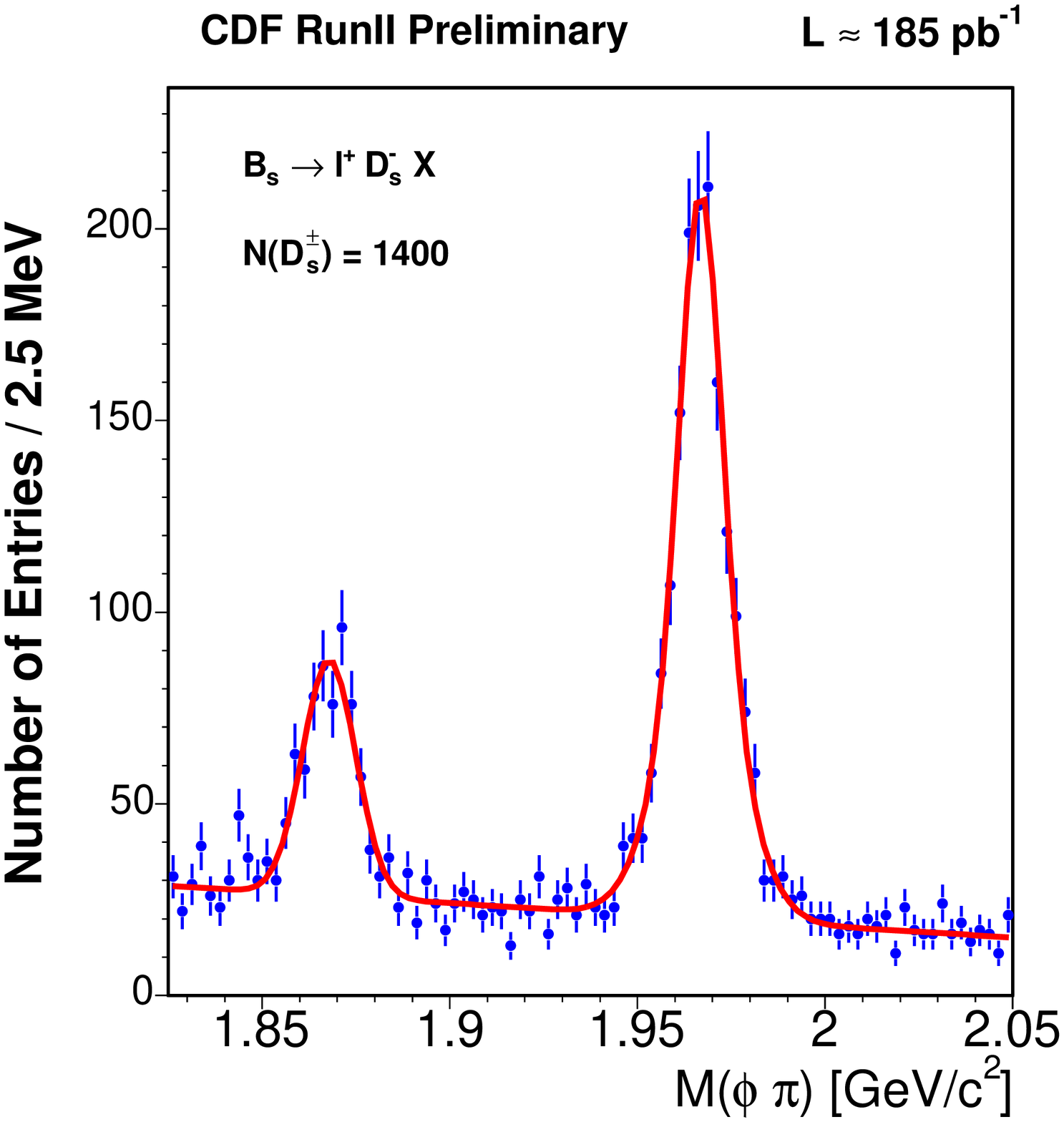}
  \caption{\it
	Left: $KK\pi$ invariant mass from 
$B_s\rightarrow D_s^-\mu^+X\rightarrow [\phi\pi^-]\mu^+X\rightarrow [[K^+K^-]\pi^-]\mu^+X$ reconstructed by D\O. Right: $KK\pi$ invariant mass in 
the same  mode reconstructed by CDF II using also electrons.
    \label{fig:bsemilept} }
\end{figure}

The \emph{left} plot in Figure~\ref{fig:bsemilept} shows the $KK\pi$ 
invariant mass
from  $B_s\rightarrow D_s^-\mu^+X\rightarrow [\phi\pi^-]\mu^+X\rightarrow [[K^+K^-]\pi^-]\mu^+X$ decays collected by D\O. The two peaks correspond
to the $D^-$ and $D_s^-$ states, which can both decay to $\phi\pi^-$.
The \emph{right} plot in Figure~\ref{fig:bsemilept} shows the same 
distribution from CDF II. D\O~triggers only on muons and has a specific yield
 of $\sim 38$ pb. Even though CDF II triggers on muons and electrons, its 
specific yield is $\sim 5$ times smaller. Such a large difference comes from
the larger D\O~acceptance, however CDF II samples are cleaner
(by requiring also a displaced track) with better (a factor of $\sim$ 2) 
mass resolution.   
\subsubsection{\it $B_s$ Mixing Samples: Exclusive Decays}
\label{sssec:bsexclusive}
\begin{figure}[t]
  \vspace{6.0cm}
  \includegraphics{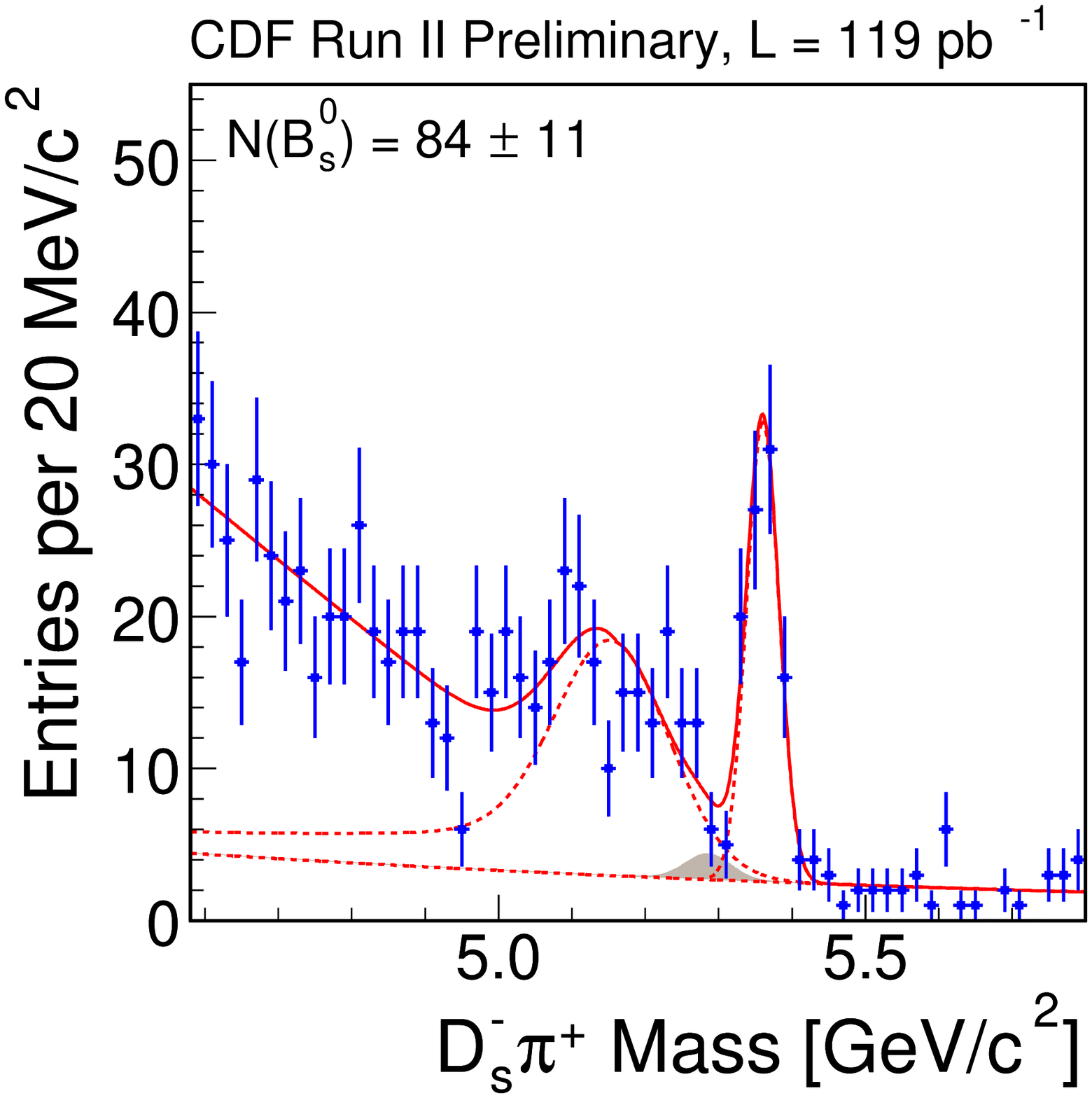}
 \includegraphics{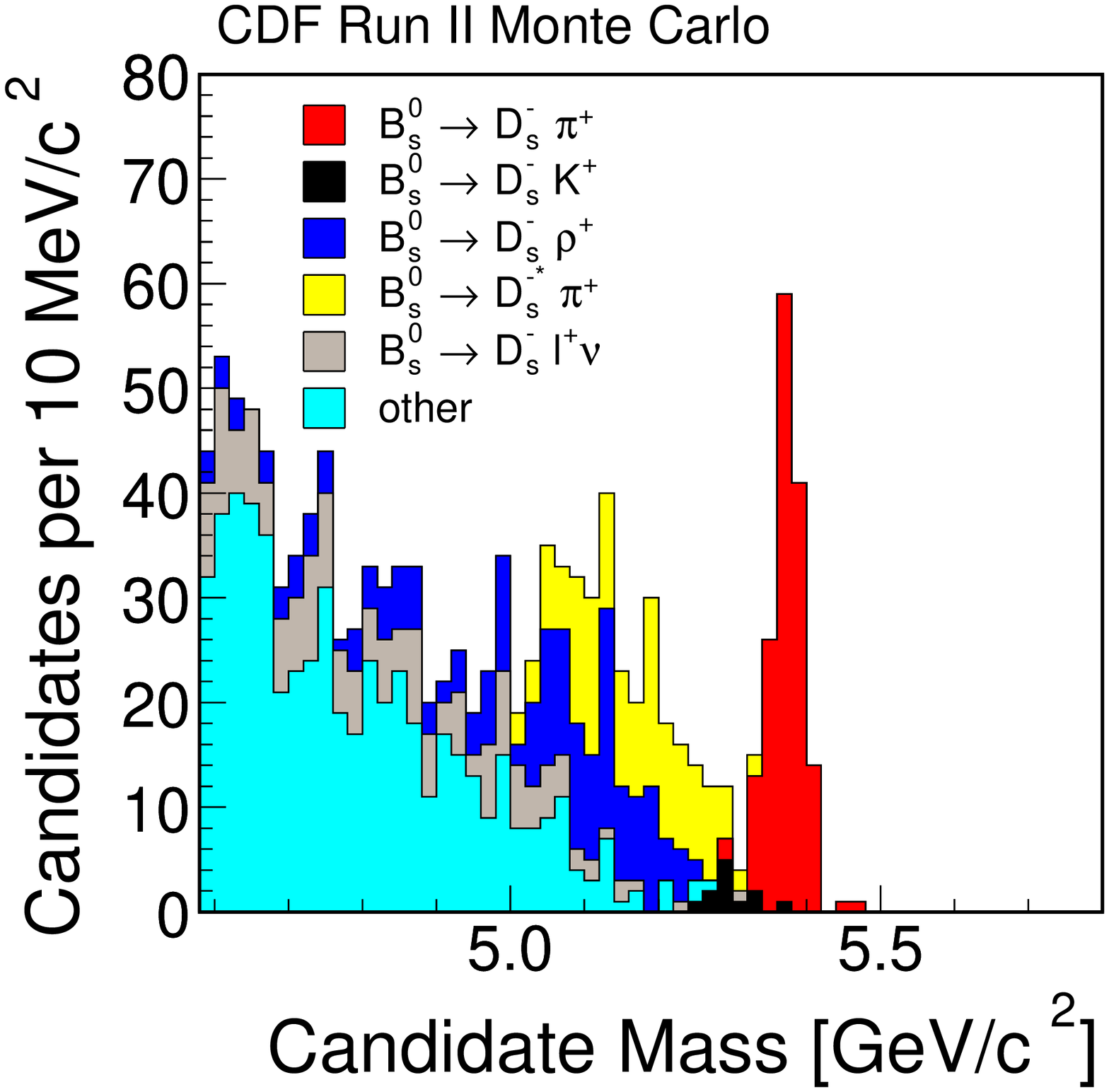}
  \caption{\it Left: $KK\pi\pi$ invariant mass from 
$B^0_s\rightarrow D^-_s\pi^+$ decays at CDF II. Right: same distribution from
Monte Carlo simulation.
    \label{fig:b_s->d_spi} }
\end{figure}

Fully reconstructed modes offer fewer signal events with better
proper time resolution and are considered the only useful ones at high values
of $\Delta m_s$. The CDF II displaced track trigger has  accumulated a sample
of exclusive  $B_s\rightarrow D^-_s\pi^+\rightarrow [\phi\pi^-]\pi^+\rightarrow [[K^+K^-]\pi^-]\pi^+$ as shown in  the \emph{left} plot in Figure~\ref{fig:b_s->d_spi}. 
A clear $B_s$ peak is visible with good purity (S/B $\sim$ 2), the broad
 shoulder at lower masses is the
 $B_s\rightarrow D_s^{*-}\pi^+$,  where the photon from the $D^*_s$ decay 
is not reconstructed. The \emph{right} plot shows the expected contributions
from a $b\bar{b}$ Monte Carlo simulation. Since the simulation provides a
 very good description of the sample, signal and sidebands in data are fit
using the shapes from the Monte Carlo with floating normalizations. As a 
result, CDF II performed the first measurement of the branching ratio for
 this mode:
\begin{equation}
\frac{f_s}{f_d}\cdot\frac{BR(B_s\rightarrow D^-_s\pi^+)}{BR(B^0\rightarrow D^-\pi^+)} = 0.35 \pm 0.05~(stat.) \pm 0.04~(syst.) \pm 0.09~(BR)
\label{eq:B_s->D_spi}
\end{equation}
where $f_s$ and $f_d$ are the fragmentation functions, and the systematic 
error deriving from the uncertainty on $BR(D^-_s\rightarrow\phi\pi^-)$ 
 is quoted separately. The result is quoted as a ratio of $BR$s' in order to cancel out  many common systematics in trigger and reconstruction efficiencies.
\subsubsection{\it $B_s$ Mixing: Prospects}
\label{sssec:bmixprospects}
In terms of $B_s$ mixing performance, D\O~takes advantages from the very high
semileptonic yields and a total higher tagging power. CDF II has, instead, 
 cleaner samples, better mass resolution and privileged 
access to larger samples of exclusive decays with better proper 
time resolution. CDF II and D\O~will extend the search  for $B_s$ mixing,
 however this measurement is
 very challenging and will  take time, effort and a significant data 
sample.\par Assuming the current performance in terms of yield, purity, 
proper time resolution and flavor tagging, CDF II estimates a $2\sigma$ 
sensitivity for $\Delta m_s = 15$ ps$^{-1}$ with about 500 pb$^{-1}$ 
of integrated luminosity (year 2005) in the exclusive modes.
 However, some improvements to the  current running configuration are
 in progress. Additional modes  both for the 
$D_s$  ($D_s^-\rightarrow K^{*}K^-, K^0_SK^-$) and for the $B_s$ 
($B_s\rightarrow D_s^-\pi^+\pi^-\pi^+$) will increase $B_s$ yields 
by $\sim20\%$.  The proper time resolution is expected to improve soon to
 $\sigma_{c\tau}\simeq$ 50 fs after  exploiting  
  the very first silicon layer (1.6 cm from the beam) and  an optimized
 calculation of the beam spot position. After optimization of all flavor
tagging algorithms, CDF II expects to reach $\epsilon D^2 = 5\%$ in flavor 
tagging performance.
With these modest improvements, 2 to 3 fb$^{-1}$
 of integrated  luminosity would be needed for CDF II to scan 
the region of $\Delta m_s$   currently preferred by indirect
 fits\cite{ciuchini}.\par
D\O~instead exploits its large semileptonic yields  
($\sim$ 30K/fb$^{-1}$ expected) and  impressive flavor tagging
 performance (expected $\epsilon D^{2}\simeq$ 10\%) and estimates to reach
 a 1.5$\sigma$ sensitivity for $\Delta m_s = 15$ ps$^{-1}$ with 500 pb$^{-1}$.
\subsubsection{\it CDF Run I: Average Time-Integrated Mixing 
Probability}
\label{sssec:giromini}
CDF measured recently the average time-integrated mixing
probability on the full Run I data sample\cite{giromini}.
 The ratio  $R=LS/OS$ of like-sign ($LS$) to opposite-sign
 ($OS$) di-leptons  was measured in  $\mathcal{L}\simeq 110$ pb$^{-1}$ of
 double semileptonic  decays of $b\bar{b}$ pairs. A two-dimensional fit of 
the lepton impact parameters in $e\mu$ and $\mu\mu$ samples selects
leptons from $b$ decays. $R$ is related to the average 
time-integrated mixing parameter:
\begin{equation}
\overline{\chi} = \frac{\Gamma(B_{d,s}\rightarrow \overline{B}_{d,s}\rightarrow l^+X)}{\Gamma(b\rightarrow l^\pm X)} = f'_d\chi_d + f'_s\chi_s
\label{eq:chibar} 
\end{equation}
where the denominator is the semileptonic width of all $b$-hadrons and
 $f'_{d}$ ($f'_{s}$) are the fragmentation functions 
$f_s$ ($f_d$) weighted with the 
corresponding semileptonic branching ratio\footnote{$f'_{q} =f_{q}/(\Gamma_q\tau_b)$ $(q= s, d)$ where $\Gamma_q$ is the semileptonic width of the $B_q$ 
meson and $\tau_b$ is the average $b$-hadron lifetime.}.
 $\overline{\chi}$ is a probe for either flavor mixing or $B$ meson
 fragmentation. The CDF result is: 
$\overline{\chi} = 0.152 \pm 0.007~(stat.) \pm 0.011~(syst.)$. This value
is higher than the current world average\cite{pdg}, 
$\overline{\chi}_{PDG} = 0.118 \pm 0.005$, that is dominated by the 
measurements at the $Z^0$ pole. 
\subsection{\textsf{CP} Violation in $B_{d(s)}\rightarrow h^+h^{'-}$ Decays}
\label{ssec:bhh}
\begin{figure}[t]
  \vspace{6.0cm}
  \includegraphics{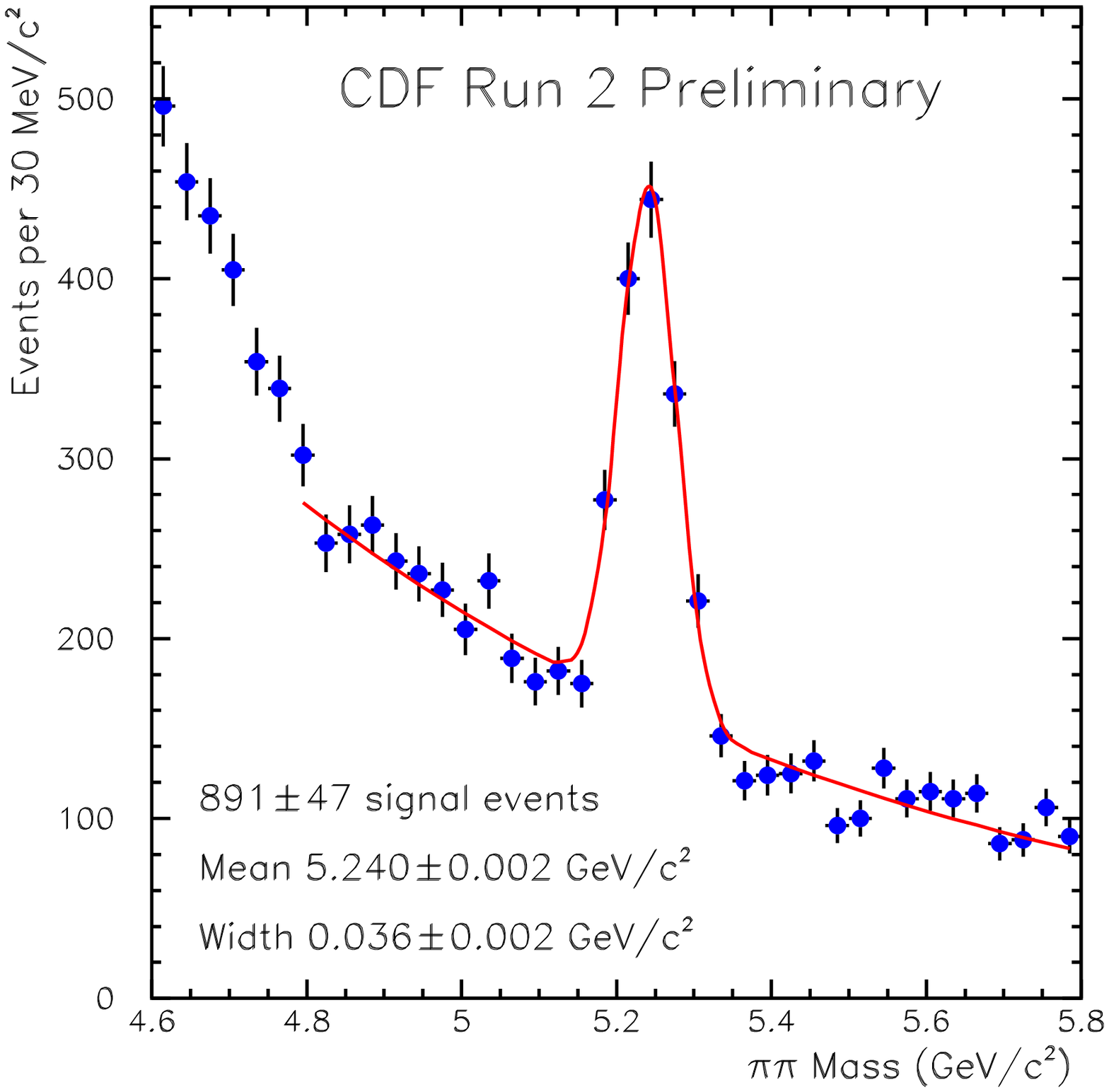}
 \includegraphics{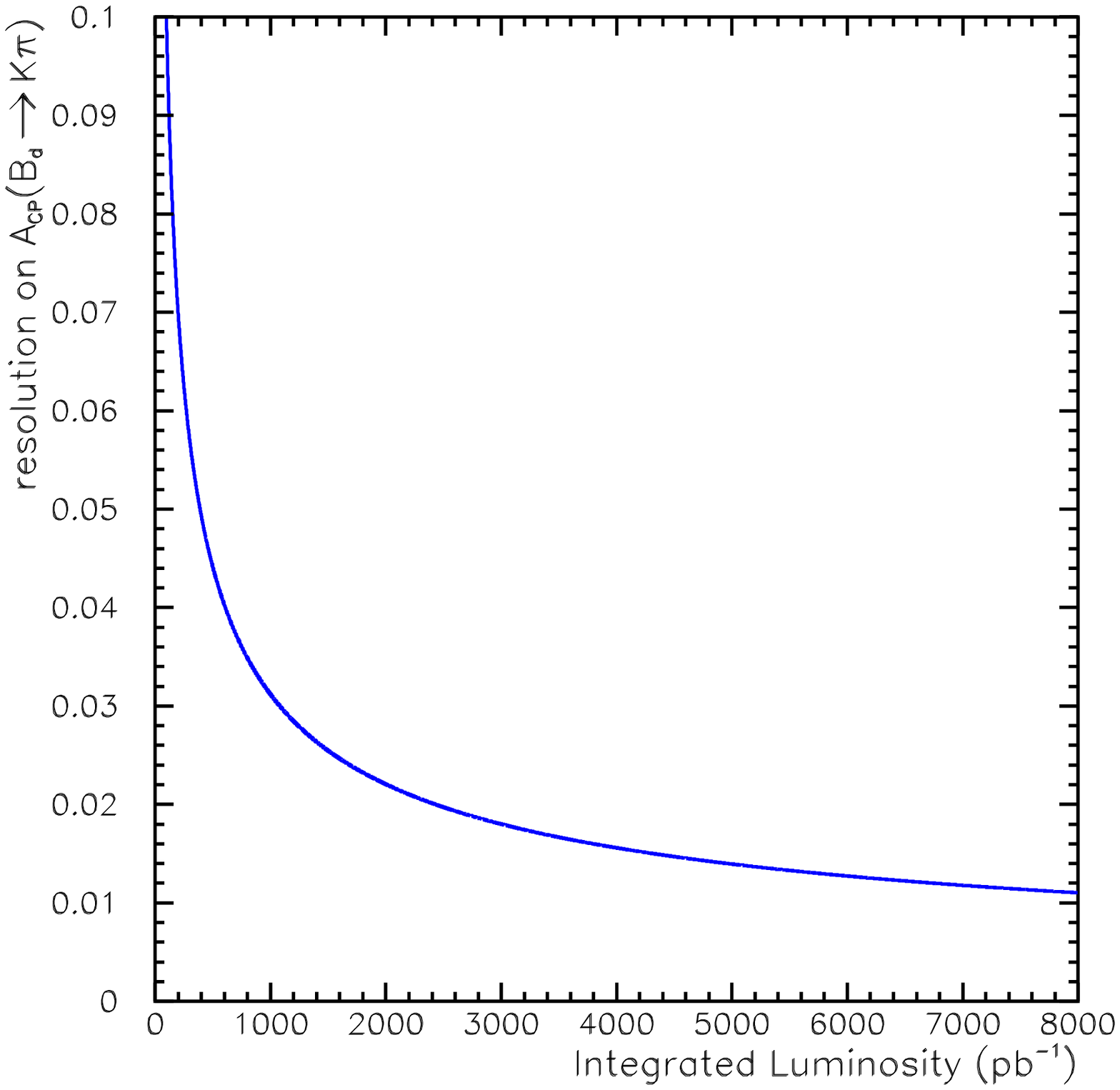}
  \caption{\it
    Left: $\pi\pi$ invariant mass spectrum of $B_{d(s)}\rightarrow h^+h^{'-}$
 candidates. Right: expected statistical resolution versus luminosity for
 the \emph{direct} \textsf{CP} asymmetry measurement in $B^0\rightarrow K^+\pi^{-}$ at CDF II.
    \label{fig:bhh1} }
\end{figure}

Using the new trigger on displaced tracks, CDF II has collected several hundred
events of charmless $B^0$ and $B_s$ decays in two tracks from 
$\mathcal{L}\simeq 180$ pb$^{-1}$ integrated luminosity. The invariant
 mass spectrum of the $B_{d(s)}\rightarrow h^+h^{'-}$ candidates
 with pion mass  assignment for both tracks is shown in the \emph{left} plot 
in Fig.~\ref{fig:bhh1}.
 A clear peak  is seen, and its width ($\sigma\simeq$ 36 MeV/c$^2$) is
 significantly  larger than the intrinsic CDF II resolution. 
This happens because (at least) four different channels overlap under
 the peak: $B^0\rightarrow \pi^+\pi^{-}$, $B^0\rightarrow K^+\pi^{-}$,
 $B_{s}\rightarrow K^+K^{-}$ $B_{s}\rightarrow \pi^+K^{-}$.
 One of the key physics goals of CDF II is to measure time-dependent
 decay \textsf{CP} asymmetries in flavor-tagged samples
of  $B^0\rightarrow \pi^+\pi^{-}$  and $B_{s}\rightarrow K^+K^{-}$
 decays. This method was first
 suggested by Fleischer\cite{fleischer} and consists of fitting simultaneously 
the four \textsf{CP} asymmetries $A_{\mathsf{CP}}^{dir-\pi\pi}$,
 $A_{\mathsf{CP}}^{mix-\pi\pi}$, $A_{\mathsf{CP}}^{dir-KK}$ and 
$A_{\mathsf{CP}}^{mix-KK}$ with:
\begin{equation}
A_{\mathsf{CP}}^{B^0}(t) = A_{\mathsf{CP}}^{dir-\pi\pi}\cos(\Delta m_dt) + A_{\mathsf{CP}}^{mix-\pi\pi}\sin(\Delta m_dt)
\label{eq:fleischerbd} 
\end{equation}
\begin{equation}
A_{\mathsf{CP}}^{B_s}(t) = A_{\mathsf{CP}}^{dir-KK}\cos(\Delta m_st) + A_{\mathsf{CP}}^{mix-KK}\sin(\Delta m_st)
\label{eq:fleischerbs} 
\end{equation}
Theoretically, one assumes U-spin symmetry\footnote{The U-spin symmetry 
is a subgroup of flavor SU(3)  that transforms the $d$ quark into an $s$ 
quark transforming thus $B^0\rightarrow\pi^+\pi^-$ into
 $B_s\rightarrow K^+K^-$.} and combines the  $B^0\rightarrow\pi^+\pi^-$
 and $B_s\rightarrow K^+K^-$ modes to cancel out the uncertainties coming
 from  hadronic  \emph{penguin}\footnote{$B^0\rightarrow\pi^+\pi^-$ and 
$B_s\rightarrow K^+K^-$ decay amplitudes receive significant contribution from 
second-order non-perturbative hadronic diagrams (\emph{i.e.} penguin diagrams). Since they carry  weak phases  which differ from those of tree-level
 diagrams, the extraction of CKM parameters from \textsf{CP}
 asymmetries becomes
 more complicated.} diagrams. This method would 
 allow a reasonably clean determination of  the CKM angle $\gamma$.
First step toward the time-dependent analysis is to disentangle the
 different contributions to the $B_{d(s)}\rightarrow h^{+}h^{'-}$ signal 
shown in the \emph{left} plot in Fig.~\ref{fig:bhh1}.
Since the TOF  $K/\pi$ separation is marginal in this momentum regime 
($p(h)> 2$ GeV/c) and the \emph{dE/dx} separation-power is limited
to $1.4\sigma$,  an event-by-event separation looks very difficult.
Therefore CDF II exploits the statistical separation provided by the
 combination  of kinematics differences between the contributing modes with the
 PID from \emph{dE/dx}. An un-binned maximum likelihood fit is performed relying
 on two discriminating variables. The first one
 is the \emph{dE/dx} information 
 calibrated on charged $K$ and $\pi$ from $D^0$ in
 about 300,000 $D^{*+}$ decays 
\footnote{See Section~\ref{sssec:giagu} for the  motivation for using $D^{*\pm}$.}. The other  variable is the
kinematic-charge correlation between the invariant mass (with pion 
assignment) $M_{\pi\pi}$, and the signed momentum imbalance between 
the two tracks. This quantity is written as 
$(1-p_{min}/p_{max})\cdot q_{min}$ where $p_{min} (p_{max})$ is the
scalar momentum of the track with the smaller (larger) momentum and
 $q_{min}$ is the charge of the track with smaller momentum.
\begin{figure}[t]
  \vspace{8.0cm}
  \includegraphics{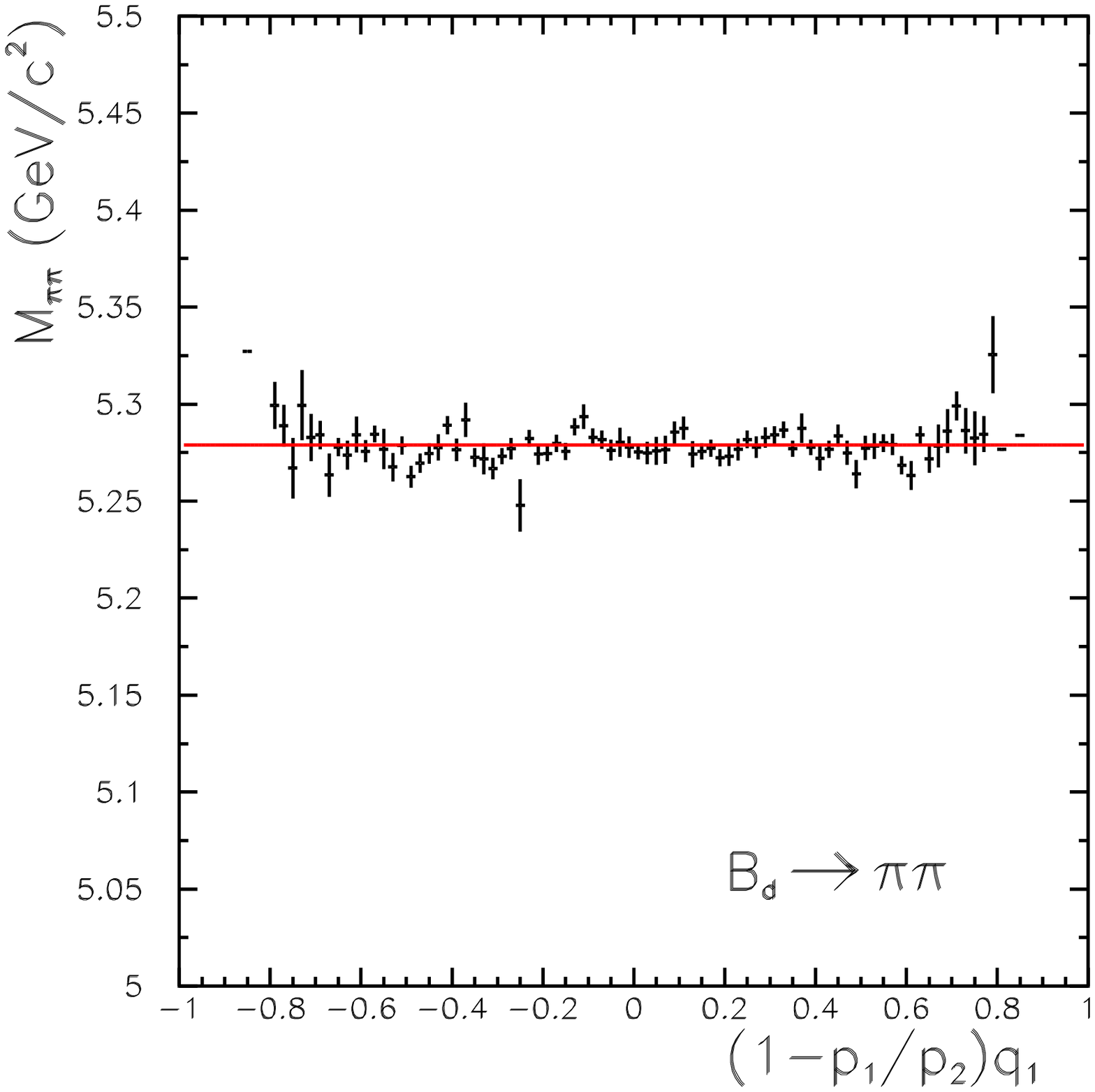}
  \includegraphics{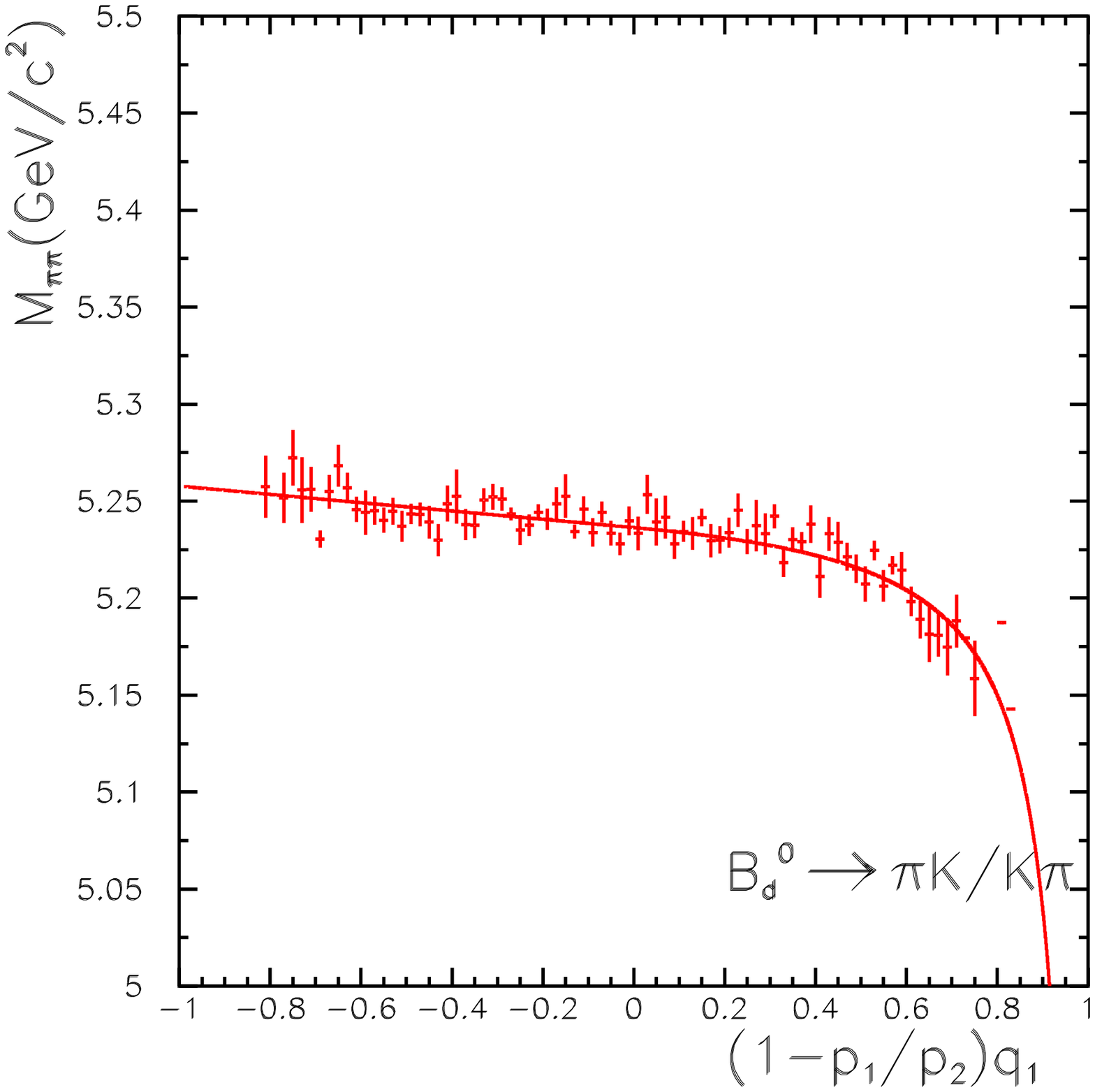}
  \includegraphics{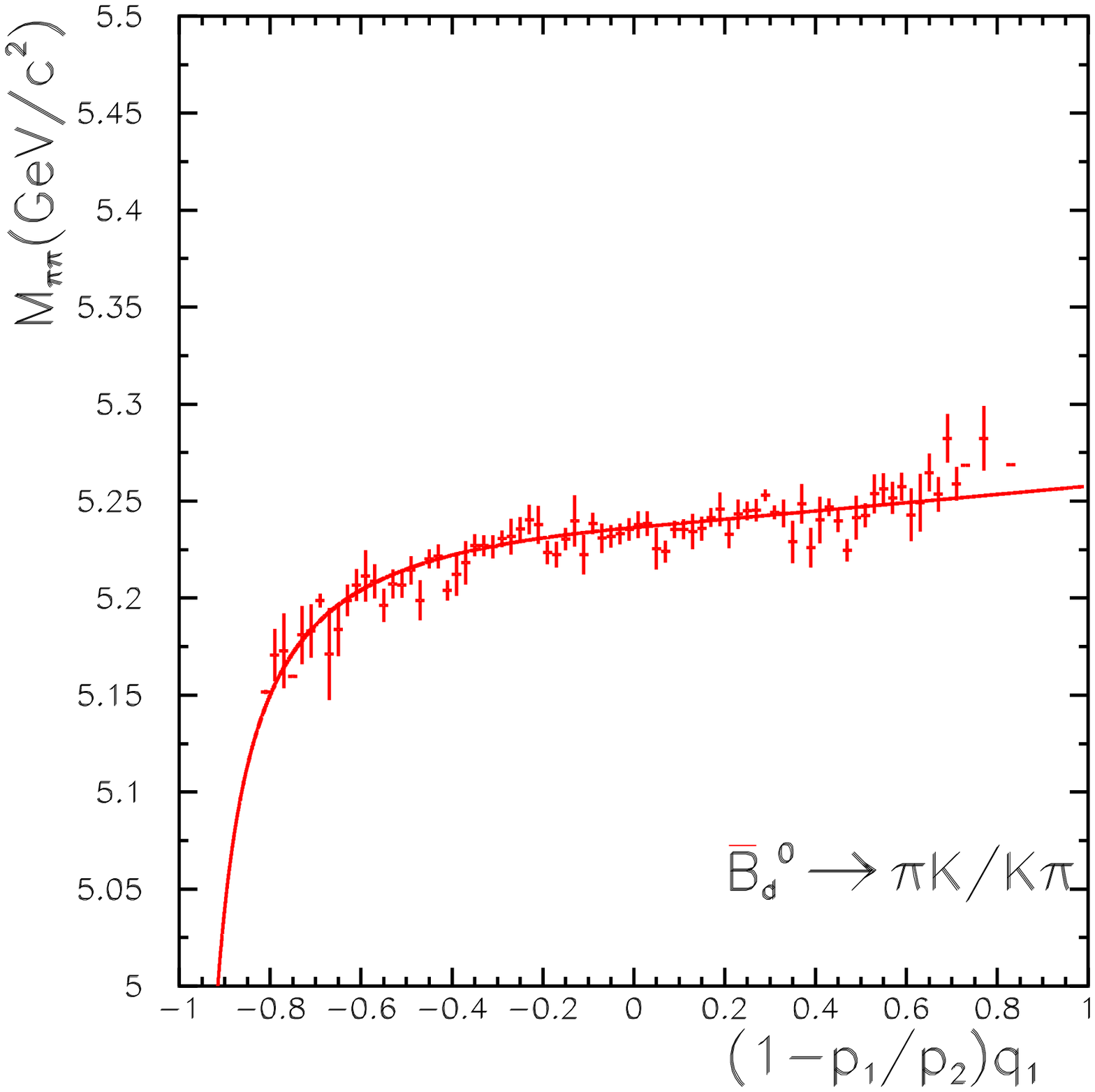}
  \includegraphics{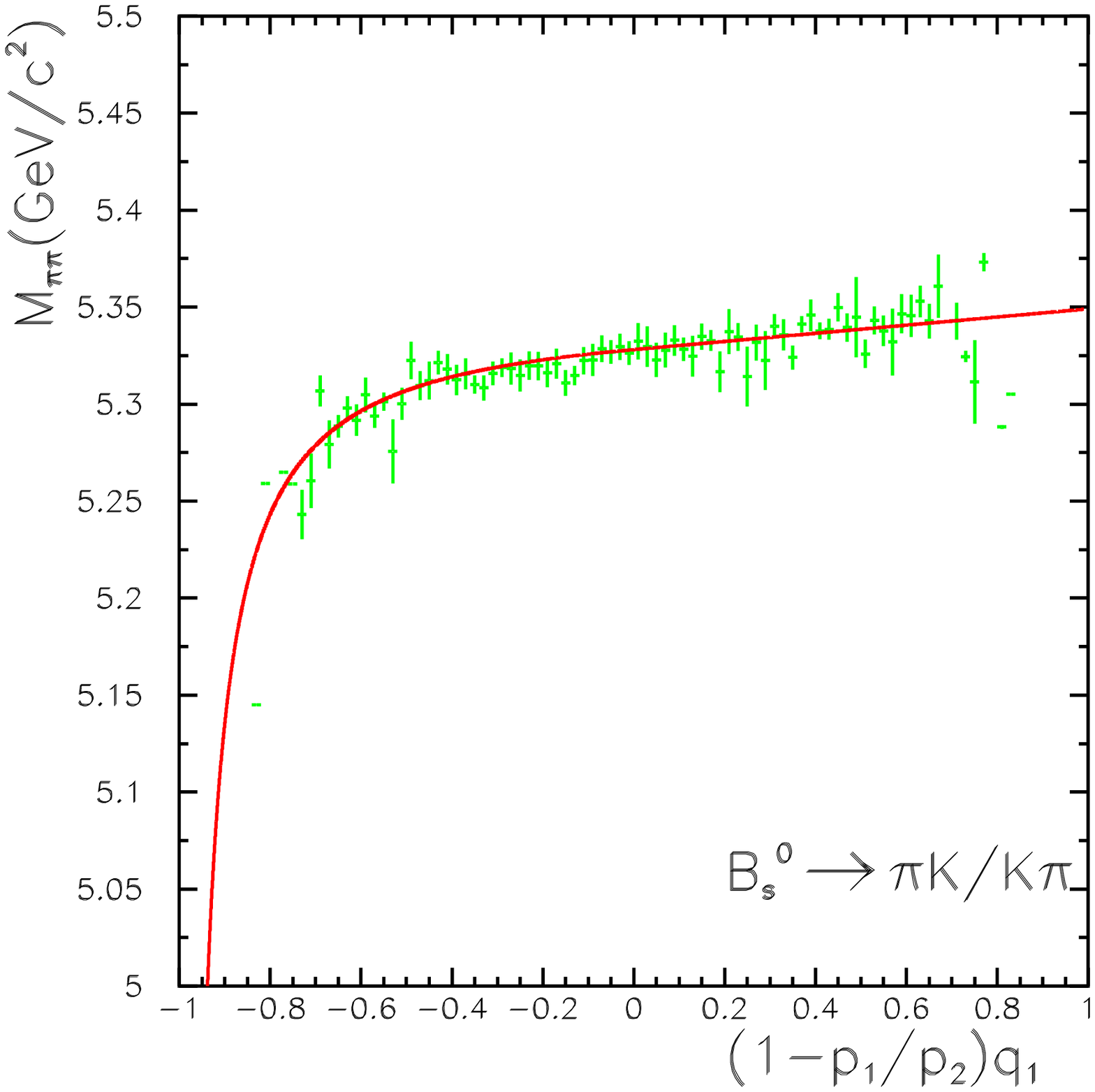}
  \includegraphics{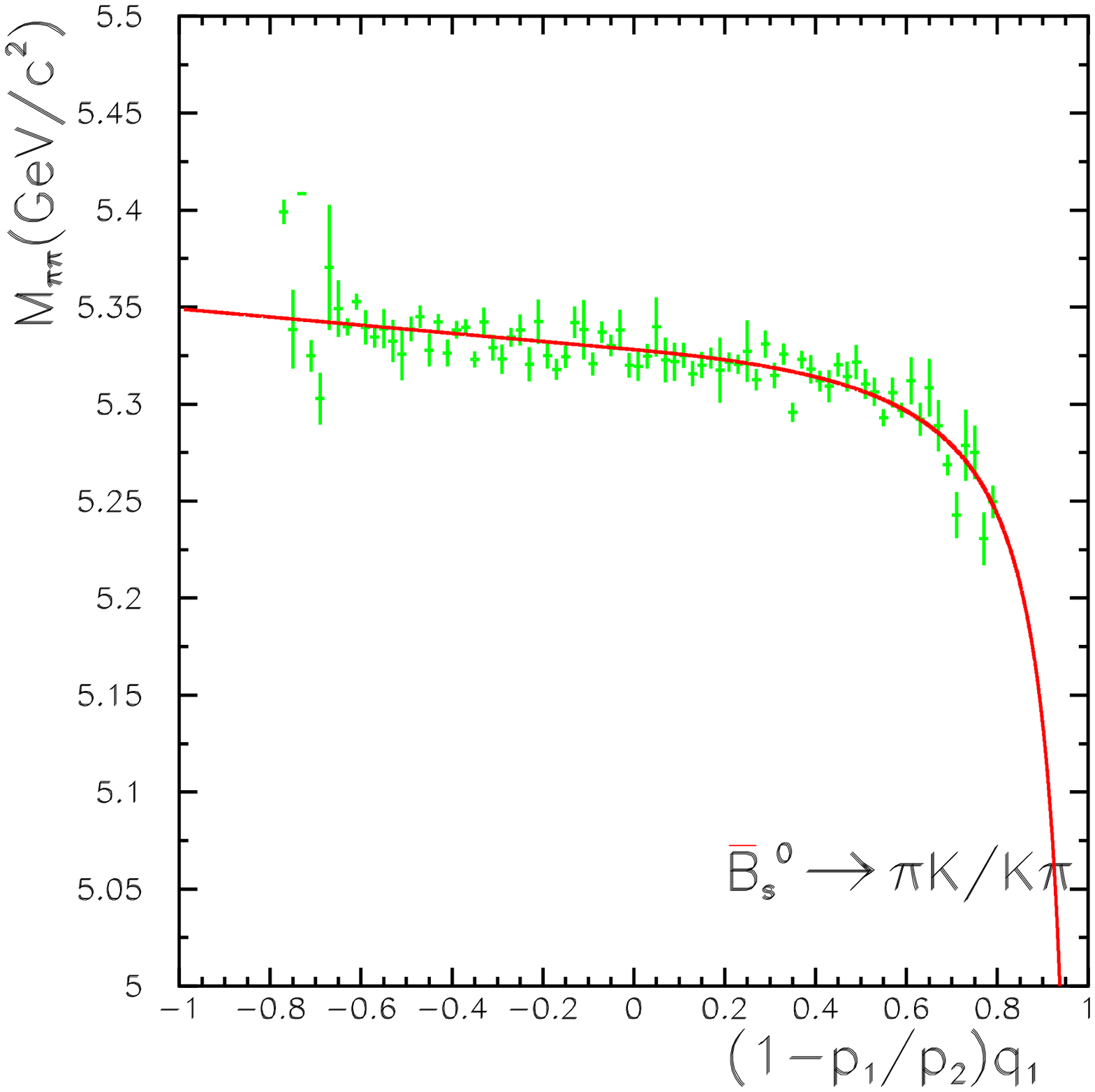}
  \includegraphics{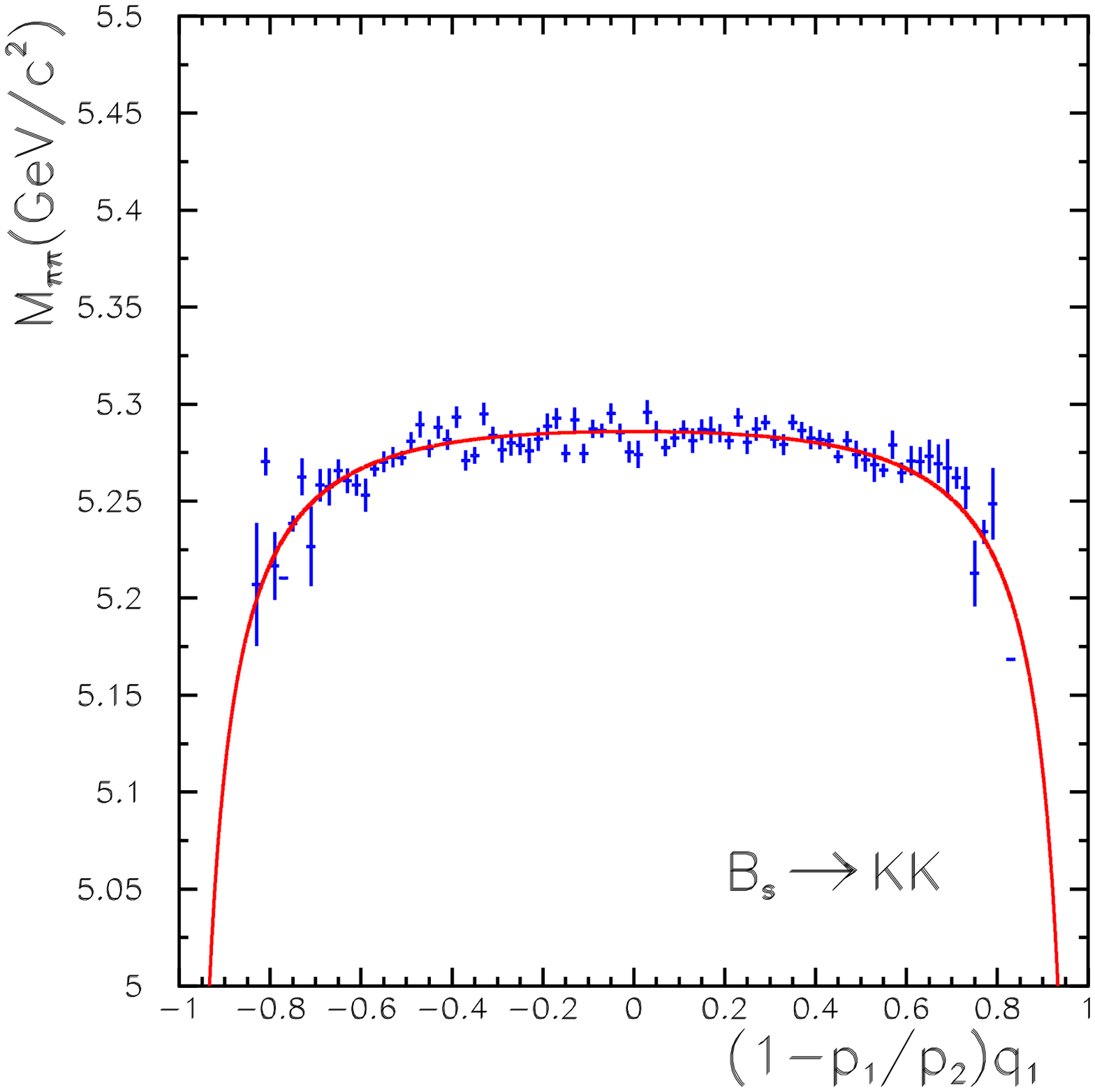}
  \caption{\it
    Monte Carlo distributions of the averaged $M_{\pi\pi}$ versus 
$(1-p_{min}/p_{max})\cdot q_{min}$ for each channel.
 Colored dots are Monte Carlo data,  solid red lines are the analytic 
functions used in the likelihood.  
    \label{fig:alpha} }
\end{figure}
 The distribution 
from a Monte Carlo simulation of $M_{\pi\pi}$ versus  
$(1-p_{min}/p_{max})\cdot q_{min}$  is shown in Fig.~\ref{fig:alpha}. The plots
show how the kinematic-charge correlation distinguishes
 also $K^+\pi^-$ from $K^-\pi^+$  final states providing \emph{direct}
 \textsf{CP} asymmetry information.\par The results from
 a subsample (65 pb$^{-1}$) of the events shown in the \emph{left} plot in 
Fig.~\ref{fig:bhh1} are summarized in the second column of 
 Table~\ref{tab:bhhyields}.
\begin{table}[t]
  \centering
  \caption{ \it CDF II results on two-body charmless B decays in 65 pb$^{-1}$ (second column), projected yields for 3.5 fb $^{-1}$ (third column).}
  \vskip 0.1 in
  \begin{tabular}{lcc}
      Mode & Fitted Yield [events] & Projected Yield/3.5 fb$^{-1}$[evts.]\\
    \hline
$B^0\rightarrow K^+\pi^{-}$&148$\pm$17 \emph{(stat.)}$\pm$17
 \emph{(syst.)} & $\sim$ 11,700\\
$B^0\rightarrow \pi^+\pi^-$&39$\pm$14 \emph{(stat.)}$\pm$17 
\emph{(syst.)}& $\sim$ 3,100\\
$B_{s}\rightarrow K^+K^{-}$  &90$\pm$17 \emph{(stat.)}$\pm$17 
\emph{(syst.)}& $\sim$ 7,100\\
$B_{s}\rightarrow \pi^+K^{-}$&3$\pm$11 \emph{(stat.)}$\pm$17 
\emph{(syst.)}& $\sim$ 1,900\\
    \hline
  \end{tabular}
  \label{tab:bhhyields}
\end{table}
The measurement presented here is the first observation of the decay $B_s\rightarrow K^+K^-$ with a relative branching fraction of:
\begin{equation}
  \frac{BR(B_s\rightarrow K^+ K^-)}{BR(B^0\rightarrow \pi^- K^+)} =
 2.71\pm 0.73~(stat.) \pm 0.35~(f_s/f_d) \pm 0.81~(syst.)
  \label{eq:bskk} 
\end{equation}
using the world average measurement of the fragmentation fraction $f_s/f_d= 0.27\pm 0.04$\cite{pdg}. The measured \emph{direct} \textsf{CP} asymmetry in the $B^0\rightarrow\pi^- K^+$ mode is:
\begin{equation}
  \frac{N(\overline{B^0}\rightarrow K^- \pi^+) - N(B^0\rightarrow \pi^- K^+)}{N(\overline{B^0}\rightarrow K^- \pi^+) + N(B^0\rightarrow \pi^- K^+)} = 0.02\pm0.15~(stat.)\pm 0.02~(syst.)
  \label{eq:bdkpiacp} 
\end{equation}
Systematic  uncertainties in all these results are dominated by the,
 still preliminary,  \emph{dE/dx} calibrations used. 
The systematic error on the \emph{direct} \textsf{CP} asymmetry is 
already comparable  to systematics in current $B$-factories measurements\cite{hfag} (Belle: 
$A_{\mathsf{CP}}^{dir}(B^0\rightarrow\pi^- K^+)= -0.088~\pm~0.035~(stat.)~\pm
~0.018~(syst)$). The  15\% statistical error of 
$A_{\mathsf{CP}}^{dir}(B^0\rightarrow\pi^- K^+)$ is a promising achievement
 considering that is
obtained in a sample of only 65 pb$^{-1}$ (the current CDF II sample is already
 three times larger).\\The yield projections\footnote{Since the CDF 
II result is not yet sensitive to the $B_s\rightarrow K^-\pi^+$,
 the projection for this mode comes from theoretical prediction
 of the branching fraction.}
 for $\mathcal{L}= 3.5$ fb$^{-1}$ are  summarized in the third column 
of Tab.~\ref{tab:bhhyields}. The \emph{right} plot in Fig.~\ref{fig:bhh1} 
 shows the expected
resolution on the \emph{direct} \textsf{CP} asymmetry in the 
$B^0\rightarrow\pi^-K^+$ mode  basing on the foreseen yields:
 CDF II will be competitive with current
 $B$-factories results with less than 1 fb$^{-1}$ of data. A projection for the
 time-dependent  analysis performed on flavor-tagged samples needs some
ingredients that are still being optimized such as the flavor-tagging 
performance
and the proper-time resolution. CDF II estimates that  data in excess of
 4 fb$^{-1}$ integrated luminosity are needed to reach
 $\mathcal{O}(20\%)$ uncertainties.  However, an intermediate goal could 
 be to extract information on $\gamma$ using just the measurements of branching ratios 
together with some minimal  dynamic assumptions, as suggested in\cite{fleischer2}.   
\subsection{\textsf{CP} Violation in Other Modes}
\subsubsection{\it \textsf{CP} Violation with Charm: $D^0\rightarrow h^+h^{'-}$ Decays}
\label{sssec:giagu}
The new CDF II trigger on displaced tracks is highly effective in collecting 
large samples of \emph{charmed} decays. Specific yields in excess of 2 nb were
 measured for $D^{*+}\rightarrow D^0\pi^{+}\rightarrow[K^-\pi^+]\pi^+$ modes
and allowed the best measurement of the \emph{direct} \textsf{CP} violating 
decay rate asymmetry of $D^0\rightarrow K^+K^-$ and 
$D^0\rightarrow \pi^+\pi^-$ to date.  Since Standard Model (SM) expectations for
 \emph{direct} \textsf{CP} violation in such modes are generally small,
 $\mathcal{O}(10^{-3})$, non-SM \textsf{CP} violation sources could appear
 if anomalously high asymmetries would be measured. CDF II uses $D^0$ from
 $D^*$ decays because (1) 
the charge of the soft pion from the $D^*$ identifies uniquely the $D^0$
 flavor, (2) a tight cut on $M(D^*)-M(D^0)$ reduces strongly the reflection
 background. An invariant mass  fit in 123 pb$^{-1}$ of data 
 reconstructs about 7,300 $D^0\rightarrow \pi^+\pi^-$ with $\sim$ 93\% purity,
 and about 16,200  $D^0\rightarrow K^+K^-$ with $\sim$ 75\% 
purity\footnote{The higher background contamination comes from partially
 reconstructed $D^0\rightarrow K^-\pi^+\pi^0$.}. No significant \emph{direct} 
\textsf{CP} violation in Cabibbo-suppressed $D^0$ decays is found: 
\begin{equation}
  A_{\mathsf{CP}}^{dir}(D^0\rightarrow K^+K^-)= [2.0\pm 1.2~(stat.)\pm 0.6~(syst.)]\%
  \label{eq:Dkk} 
\end{equation}
\begin{equation}
  A_{\mathsf{CP}}^{dir}(D^0\rightarrow \pi^+\pi^-)= [1.0\pm 1.3~(stat.)\pm 0.6~(syst.)]\%
  \label{eq:Dpipi} 
\end{equation}
The statistics on the control sample, used to measure residual effects on the 
intrinsic detector charge asymmetry, dominates the systematics uncertainty.
\subsubsection{\it \textsf{CP} Studies with $B^0_s\rightarrow J/\psi\phi$ Decays}
\label{sssec:Bsjpsiphi}
\begin{figure}[t]
  \vspace{4.3cm}
  \includegraphics{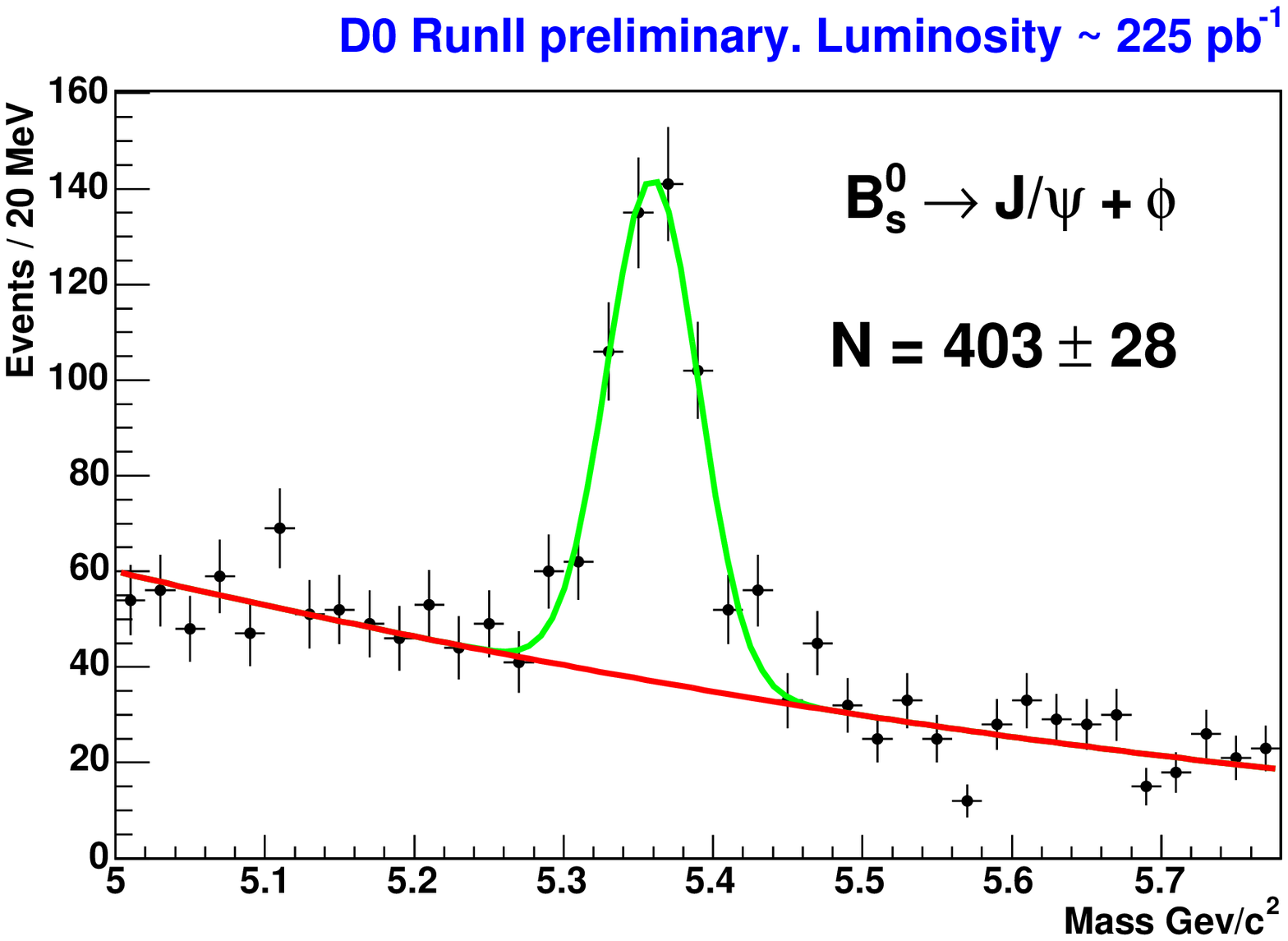}
 \includegraphics{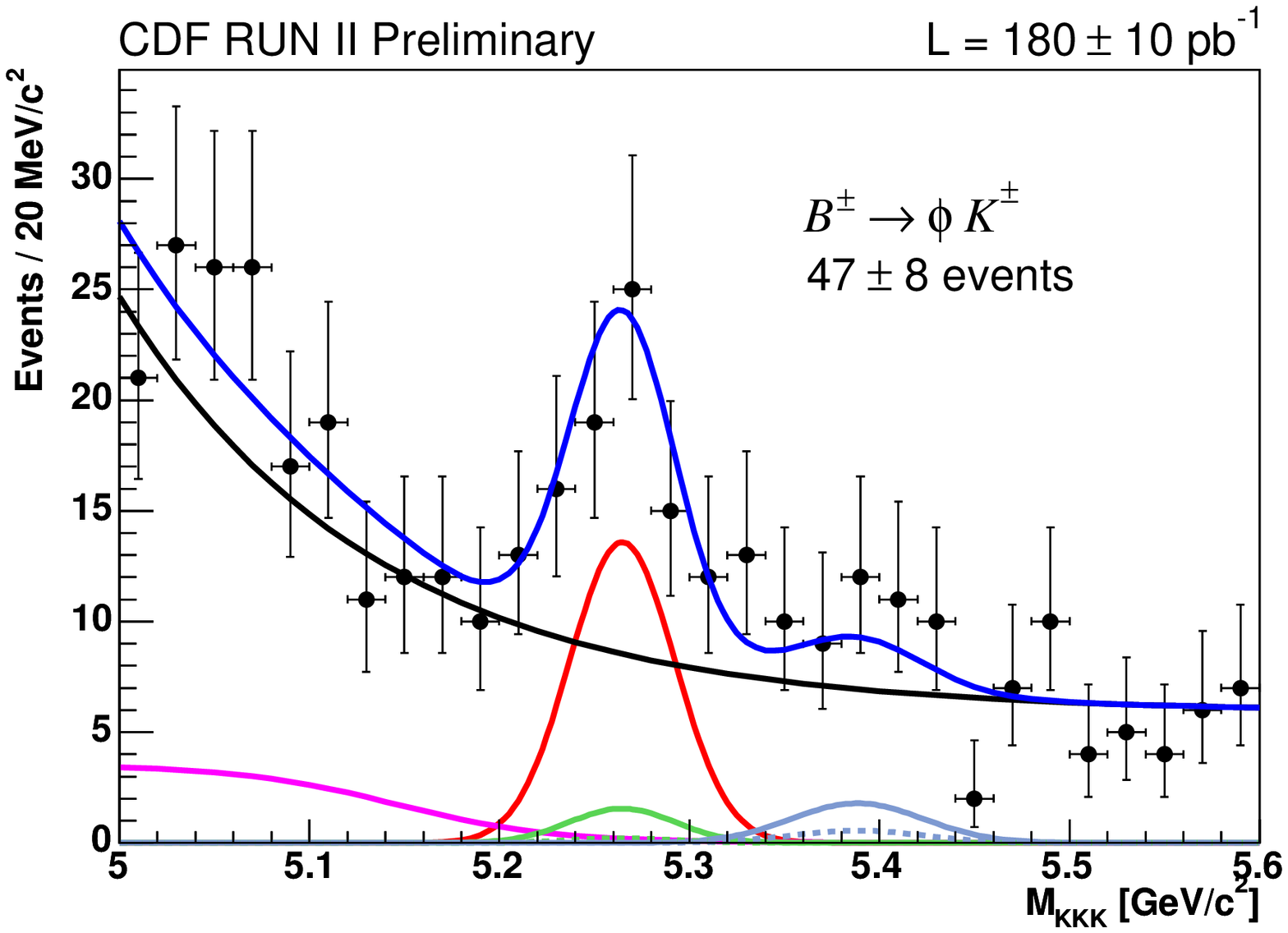}
  \caption{\it
 Left: $KK\mu\mu$ invariant mass in $B^0_s\rightarrow J/\psi\phi$ decays from
 D\O. Right: $KKK$ invariant mass in $B^+\rightarrow\phi K^{+}$ from CDF II.
    \label{fig:others} }
\end{figure}

$B_s\rightarrow J/\psi\phi$ decays  will be used to measure  the relative
 lifetime difference  $\Delta\Gamma_s/\Gamma_s$ between the two
  $B_s$ \textsf{CP} eigenstates. Since SM predicts the ratio 
$\Delta\Gamma_s/\Delta m_s = \mathcal{O}(10^{-3})$,
 $\Delta\Gamma_s/\Gamma_s$ could be  a complementary method for
 discovering $B_s$ oscillations with large $\Delta m_s$ values.
 In addition, once $B_s$ oscillations will 
be established,  the time dependent decay \textsf{CP} asymmetry will
 provide information  about the $V_{ts}$  weak phase 
$\beta_s$. This number  is expected 
small  in the SM, so a significantly  large  
asymmetry would hint at New Physics. However, 
since the final states have two vector mesons,  the 
$B^0_s\rightarrow J/\psi\phi$ \textsf{CP}-parity depends on their relative 
angular momentum. Angular analysis  is required to separate \textsf{CP}-even
 from \textsf{CP}-odd decays. Thanks to the di-muon trigger
 both CDF II and D\O~reconstruct the exclusive mode  
$B^0_s\rightarrow J/\psi\phi\rightarrow[\mu^+\mu^-][K^+K^-]$.
 The \emph{left} plot in  Figure~\ref{fig:others}
 shows the D\O~invariant mass plot with 403 $\pm$ 28
 candidates in $\sim$ 225 pb$^{-1}$ of data.
 D\O~takes advantages of a better muon
 coverage achieving a larger yield than CDF II (120 $\pm$ 13 in about 
140 pb$^{-1}$)  although CDF II has higher mass resolution.
\subsubsection{\it Other Direct \textsf{CP} Asymmetries}
\label{sssec:diracp}
Samples with flavor-specific final states are used in searches for 
\emph{direct} \textsf{CP} violation. Several classes of decays are of particular 
interest. Various $B_{d(u)}\rightarrow\phi X$ decays 
($B^+\rightarrow\phi K^+$,
$B^0\rightarrow\phi K^{*0}$ and $B^+\rightarrow\phi K^{*+}$)
 are intriguing because of an apparent 
 $\sim 3.5\sigma$ disagreement between the measurements of $\sin(2\beta)$ in 
$B^0\rightarrow\phi K^0_s$ and $B^0\rightarrow J/\psi K^0_s$\cite{belle}.
 Both CDF II and D\O~will be able to 
reconstruct the above decay modes.\par CDF II has reconstructed $47\pm 8$ of the \emph{penguin}-dominated 
$B^+\rightarrow\phi K^+$ decay in $\sim$ 180 pb$^{-1}$ of data
 (see \emph{right} plot in Figure~\ref{fig:others}). Using a
 multi-dimensional likelihood fit that includes invariant mass,
 $\phi$ helicity 
and  \emph{dE/dx} information,  CDF II measured the \emph{direct}
 \textsf{CP} asymmetry in this sample. The SM expectations prescribe
 zero asymmetry in this channel. The CDF II result is:\\
\begin{equation}
  A_{\mathsf{CP}}^{dir}(B^+\rightarrow \phi K^+)= -0.07\pm 0.17~(stat.) ^{+0.06}_{-0.05} ~(syst.)
  \label{eq:phik} 
\end{equation}
This result is already competitive with the current best measurements from 
$B$-factories\cite{phik} (see Babar:
 $A_{\mathsf{CP}}^{dir}(B^+\rightarrow \phi K^+)= 0.04~\pm~0.09~(stat.)~\pm~0.01~(syst.)$ for example) despite it was obtained using only 180 pb$^{-1}$ of data. 
This  remarkable achievement is   promising  and  will be  improved soon as
 the statistic increases. 
\subsubsection{\it \textsf{CP} Studies with $B_{d(s)}\rightarrow D_{(s)}K$ Decays}
\label{sssec:B->DK}
Several methods to use $B_{d(s)}\rightarrow D_{(s)}K$ decays for a 
theoretically clean determination of the CKM angle $\gamma$ were proposed
 (see\cite{bdk} for example). However these  modes need 
strong particle identification capabilities to identify the small
 ($\sim 8\%$)  contribution of the  Cabibbo-suppressed $D_{(s)}K$
 final state among the favored $D_{(s)}\pi$. 
CDF II, thanks to the displaced track
 trigger,  has already reconstructed $B^+\rightarrow\overline{D^0}\pi^+$, 
$B^0\rightarrow\overline{D}^{(*)-}\pi^+$  and $B_s\rightarrow D_s^-\pi^+$ 
(shown in the \emph{left} plot in Fig.~\ref{fig:b_s->d_spi})
 with fairly good purity.
 Based on current yields, CDF II expects to collect $\sim 2200$ 
$B^+\rightarrow\overline{D^0}K^+$ and more than 100 $B_s\rightarrow D_sK^+$
 per fb$^{-1}$ of integrated luminosity. The extraction of information
 on $\gamma$ is a long term and challenging task
 that will require a considerable fraction of the expected Run II statistics
 and fine-tuned particle identification tools.
\section{Summary}
In the next few years CDF II and D\O~will play a key role in \textsf{CP}
 studies using  \emph{charmed} and \emph{beauty} decays.
 The broad physics program includes both measurements competitive with 
$B$-factories and measurements accessible only to the Tevatron such as 
$B_s$ mixing and $b$-baryons studies.\par The understanding of low-level tools
 such as tracking and \emph{dE/dx} is excellent in both experiments.
 CDF II focused so far on the measurement of \emph{direct}
CP asymmetries in self-tagging modes where world-class results have already 
been achieved. The state of the tools for second generation analyses  is
 advanced and lead already to good performances on flavor-tagging and
 proper time resolutions. Both experiments are expected to provide 
significant contributions in the determination of the mixing parameter 
$\Delta m_s$. D\O~will exploit its higher flavor-tagging power and 
large semileptonic yields. CDF II has larger yields and better time
 resolution in the exclusive $B_s$ modes. Information on the angle $\gamma$
will be extracted at CDF II from $B_{d(s)}\rightarrow h^+h^{'-}$ decays 
collected, for the first time in an hadronic collider, by the trigger on
displaced tracks.
\section{Acknowledgments}
I would like to thank the \emph{``La Thuile 2004''} organizers for 
the opportunity to speak at this Conference and the colleagues from the
 CDF II and D\O~Collaborations  for assisting me while preparing this talk 
and this document.  
\end{document}